\documentclass[aps,prb,reprint,superscriptaddress,twocolumn,10pt]{revtex4-2}

\usepackage{graphicx}   
\usepackage{wrapfig}    
\usepackage{float}      
\usepackage{bm}
\usepackage[usenames,dvipsnames,svgnames,table]{xcolor}
\usepackage{epsfig}
\usepackage{amsmath}

\usepackage{tikz}
\usetikzlibrary{calc} 
\usepackage{pgfplots}
\usepackage{soul}

\usepackage{amssymb}
\usepackage{textgreek}
\usepackage{array}
\usepackage{nccmath}
\usepackage{dsfont}
\usepackage{array}
\usepackage{cancel}
\usepackage{comment}
\usepackage{hyperref}
\usepackage{cleveref}
\crefname{figure}{Fig.}{Figs.}
\hypersetup{colorlinks=true,breaklinks,linkcolor=blue,urlcolor=blue,citecolor=blue}
\usepackage{orcidlink}

\allowdisplaybreaks

\newcommand{\kh}[1]{{\color{olive}{#1}}}

\newcommand*{\ltemp}{\multicolumn{1}{c|}{}}
\newcommand*{\rtemp}{\multicolumn{1}{|c}{}}

\begin{document}
\title{Two-site entanglement in the two-dimensional Hubbard model}
\author{Frederic Bippus \orcidlink{0009-0006-4316-6547}}
\affiliation{%
    Institute of Solid State Physics, TU Wien, 1040 Vienna, Austria}%

\author{Anna Kauch \orcidlink{0000-0002-7669-0090}}
\affiliation{%
    Institute of Solid State Physics, TU Wien, 1040 Vienna, Austria}%

\author{Gerg\H o Ro\'osz \orcidlink{0000-0002-9541-119X}} 
\affiliation{%
    HUN-REN Wigner Research Center for Physics, H-1525 Budapest, P.O.Box 49, Hungary}%

\author{Christian Mayrhofer}
\affiliation{%
    Institute of Solid State Physics, TU Wien, 1040 Vienna, Austria}%

\author{Fakher Assaad \orcidlink{0000-0002-3302-9243}} 
\affiliation{%
    Institut für Theoretische Physik und Astrophysik, Universität W\"urzburg, 97074 W\"urzburg Germany 
 }

 \affiliation{%
 W\"urzburg-Dresden Cluster of Excellence ct.qmat, Am Hubland, 97074 W\"urzburg, Germany}%
 
\author{Karsten Held \orcidlink{0000-0001-5984-8549}}
\affiliation{%
    Institute of Solid State Physics, TU Wien, 1040 Vienna, Austria}%

\date{\today}
    
\begin{abstract}{The study of entanglement in strongly correlated electron systems typically requires knowledge of the reduced density matrix. Here, we apply the parquet dynamical vertex approximation to study the two-site reduced density matrix 
at varying distance, in the Hubbard model at weak coupling. This allows us to investigate the spatial structure of entanglement in dependence of interaction strength, electron filling, and temperature.
We compare results from different entanglement measures, and benchmark against quantum Monte Carlo. }

\end{abstract}
\maketitle

\section{Introduction}
Entanglement is one of the key manifestations of quantum mechanics and builds on correlations beyond the limits of what is classically possible \cite{bell_einstein_1964}. Despite its importance, quantifying and characterizing entanglement in strongly correlated electron systems remains a significant challenge, especially at finite temperatures.
Here, the fundamental model is the Hubbard model which remarkably captures many essential features of strongly correlated electron systems, including superconductivity in cuprates \cite{Halboth2000,Honerkamp2001,Gull2015} and nickelates \cite{Kitatani2020}, pseudogaps \cite{Sakai2009, Sordi2012,simkovic_origin_2024},  kinks \cite{Held2013,Byczuk2007}, waterfalls \cite{krsnik_local_2025} and the Mott transition \cite{Gebhard1997,Georges1992a}. 

To this date, various approaches have been applied to study entanglement in the Hubbard model, yet little is known about the spatial structure of entanglement. 
On the one hand, multipartite entanglement probes such as the quantum Fisher information have proven to be valuable, because they offer experimental accessibility via susceptibility measurements \cite{hyllus_fisher_2012,hauke_measuring_2016,
mazza_quantum_2024,
laurell_witnessing_2024,
frerot_quantum_2016}.
In particular, multipartite entanglement has been detected in the pseudogap~\cite{bippus2025entanglementpseudogapregimecuprate} and strange metal regime~\cite{balut_quantum_2024}. However, the quantum Fisher information characterizes the entanglement depth of a system, offering no insight into the spatial structure of entanglement.

On the other hand, bi-partite entanglement in the pseudogap has been detected based on CDMFT~\cite{bellomia_quantum_2024}, a cluster extension to the dynamical mean field theory (DMFT). Both DMFT and CDMFT have also been applied to study entanglement at the Mott transition  \cite{bellomia_quantum_2024,bera_dynamical_2024,bellomia_quasilocal_2024,walsh_local_2019,walsh_entanglement_2020,Majumder_2025}. 
CMDFT and similar calculations with exact diagonalization (ED) \cite{roosz_two-site_2024, Rohshap_2025_QTTD}  are however limited to small clusters and thus yield only little information on the spatial structure of entanglement.

Similarly limited to small two-dimensional systems, and by entanglement itself  \cite{gauvin-ndiaye_mott_2023}, are matrix product state (MPS) methods. Nevertheless, they have provided an in-depth analysis of momentum space entanglement in the ground state of the two-dimensional Hubbard model  \cite{ehlers_entanglement_2015}.

\begin{figure}[t]
    \includegraphics[scale=1]{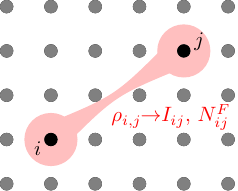}
\caption{Tracing out all but two lattice sites $i,j$ provides the two-site reduced density matrix $\rho_{i,j}$, 
computed here from two- and four-point Green's functions with the $p$D\textGamma A. 
From $\rho_{i,j}$, entanglement measures such as the mutual information $I$  and fermionic negativity $N^F$ between any two lattice sites can be evaluated. This entanglement includes spin and charge degrees of freedom.}
\label{Fig_explanation}
\end{figure}

Besides these examples, entanglement has been studied quite intensively in the one-dimensional Hubbard model \cite{roosz_two-site_2024, Rohshap_2025_QTTD,gu_entanglement_2004,abaach_long_2023,lo_schiavo_quantum_2023,wang_entanglement_2025,Consiglio_2025}. Additionally,  logarithmic corrections to the area law have been detected on the honeycomb lattice \cite{demidio_universal_2024}, and the entanglement spectra of exact excited eigenstates have been analyzed  \cite{vafek_entanglement_2017}. 
On the practical side, the study of entanglement  can be relevant, among others, to detect phase transitions \cite{DeChiara_2018,yamashika_quantum_2025} and in quantum metrology \cite{DeChiara_2018}. 
 {In general, entanglement offers a new, augmented perspective for understanding solid state phenomena.}
 {Turning to the two-site approach in particular, which is at the focus of this paper, it has been used to investigate ground state phase transitions and crossovers \cite{Rohshap_2025_QTTD}, can provide valuable insights for the optimization of MPS based numerical algorithms \cite{barcza_quantum-information_2011}. Finally, for spin systems, it has been shown that entanglement between pairs of lattice sites can be extracted and used for quantum information processing \cite{chiara_scheme_2006}.}

In this paper, we aim at studying the spatial structure of entanglement in the two-dimensional Hubbard model at finite temperatures. To this end, 
we calculate the two- and four-point Green's functions 
with the parquet dynamical vertex approximation ($p$D\textGamma A)~\cite{toschi_dynamical_2007,Valli2015,rohringer_diagrammatic_2018}.
From these we obtain,  with the methodology developed in Ref.~\onlinecite{roosz_two-site_2024},
the  two-site reduced density matrix (2s-RDM) $\rho_{i,j}$ for two lattice sites $i,j$ at arbitrary distance, see Fig.~\ref{Fig_explanation}.
The 2s-RDM allows us in turn to calculate bi-partite entanglement measures such as the mutual information $I_{ij}$ \cite{zurek_information_1983,barnett_entropy_1989} and the (fermionic) negativity $N^F_{ij}$ \cite{peres_separability_1996,horodecki_separability_1996, shapourian_entanglement_2019, shapourian_partial_2017}.


Unlike the case of ED calculations in Ref.~\onlinecite{roosz_two-site_2024}, we cannot benchmark the results against a direct evaluation of the 2s-RDM, since we now consider significantly larger systems. To provide nevertheless an independent validation, we benchmark our results for one of the entanglement measures, the Rényi mutual information, against auxiliary field quantum Monte Carlo (QMC)   \cite{Blankenbecler81,White89,Assaad08_rev,assaad_alf_2022,grover_entanglement_2013,Assaad13a}. 
For the Hubbard model on a square lattice with weak to intermediate interaction strength $U$, we study how entanglement changes with $U$, filling $n$, and inverse temperature $\beta$.
We find a close relation between entanglement and nesting of the Fermi surface.
In contrast to entanglement in pure spin systems, our approach features entanglement between charge and spin degrees of freedom.

The outline of the paper is as follows: In  Section~\ref{Entanglement_measures} we introduce the von Neumann and Rényi mutual information as well as the fermionic negativity. In Section~\ref{Model_and_Method}  the Hubbard model and the formalism to compute the 2s-RDM is recapitulated.
We then apply this post-processing routine and benchmark it against QMC in Section~\ref{QMC_comparison_main}.
In Section~\ref{Results}, the entanglement in the Hubbard model as function of distance is analyzed.  {Furthermore, in Section~\ref{Section_COR_DIAG} we analyze which correlation functions are dominant in the system to provide a physical interpretation of our findings. We dub this  {approach ``}correlation diagnostics {''}.}  
We summarize our findings and discuss future research applications in Section~\ref{Summary}.

\section{Entanglement Measures} \label{Entanglement_measures}
 {
The central quantity considered in this work is the 2s-RDM, which can be computed between any pair of lattice sites (see Fig.~\ref{Fig_explanation}). Following Ref.~\onlinecite{roosz_two-site_2024}, the matrix elements of the 2s-RDM are given by  
\begin{equation}
    \rho_{nm}=\langle A_{nm} \rangle ,
\end{equation}
with $n,m\in[1,16]$ and
\begin{equation}
    A_{n,m}=\frac{1}{\sqrt{2+2\delta_{nm}}} ( |v_n  \rangle \langle  v_m| + | v_m \rangle \langle v_n | ).
\end{equation}
\begin{table}[tb]
	\begin{center}
		\begin{tabular}{c |c c | c c| } 
             & \multicolumn{2}{c|}{$\psi_i$} & \multicolumn{2}{c|}{$\psi_j$} \\ \cline{2-5}
			 & i$\uparrow$ & i$\downarrow$  & j$\uparrow$  & j$\downarrow$  \\ \hline
		   $v_1$ & 0 & 0  & 0  & 0 \\ \hline
		   $v_2$ & 0 & 0  & 0  & 1 \\
		   $v_3$ & 0 & 0  & 1  & 0 \\
		   $v_4$ & 0 & 1  & 0  & 0 \\
		   $v_5$ & 1 & 0  & 0  & 0 \\ \hline
		   $v_6$ & 0 & 0  & 1  & 1 \\
		   $v_7$ & 0 & 1  & 0  & 1 \\
		   $v_8$ & 1 & 0  & 0  & 1 \\
		   $v_9$ &  0 & 1  & 1  & 0 \\
		   $v_{10}$ & 1 & 0  & 1  & 0 \\
		   $v_{11}$ & 1 & 1  & 0  & 0 \\ \hline
		   $v_{12}$ & 0 & 1  & 1  & 1 \\
		   $v_{13}$ & 1 & 0  & 1  & 1 \\
		   $v_{14}$ & 1 & 1  & 0  & 1 \\
		   $v_{15}$ & 1 & 1  & 1  & 0 \\ \hline
		   $v_{16}$ & 1 & 1  & 1  & 1 
		\end{tabular}
	\end{center}
		\caption{ {Occupation number basis of all two-site states $v$, expressed in terms of the 
        occupations of the two sites $i$ and $j$ and two spin species $\uparrow$ and $\downarrow$; $\psi_i$ and   $\psi_j$  denote the basis states for site $i$ and $j$ from which  the product state $v$ is made up.  Horizontal lines separate subspaces with different occupation numbers $n$. }}
		\label{tab:check}
\end{table}
The corresponding basis vectors $|v_n  \rangle$  are defined in Table~\ref{tab:check}; and only real-valued elements $A_{nm}$ exist \cite{footnote_typo}. Due to particle number and spin conservation, most matrix elements are zero, and we obtain $\rho_{ij}$ in particle number block-diagonal form}
\begin{widetext}
    \begin{equation}
        \rho_{ij} = \left(\begin{array}{cccccccccccccccc}
		\rho_{1,1} & \rtemp & &  & & & & & & & & & & & & \\ \cline{1-5}
		 \ltemp & \rho_{2,2} &  & \rho_{2,4} &  & \rtemp & & & & & & & & & &   \\
		 \ltemp &  & \rho_{3,3} &  & \rho_{3,5} & \rtemp & & & & & & & & & &   \\
		 \ltemp & \rho_{2,4} &  &  \rho_{4,4}&  & \rtemp & & & & & & & & & &   \\
		 \ltemp &  & \rho_{3,5} &  & \rho_{5,5} & \rtemp & & & & & & & & & &   \\ \cline{2-11}
		  &  &  &  & \ltemp & \rho_{6,6} &  & \rho_{6,8} & \rho_{6,9} &  & \rho_{6,11} & \rtemp  & & & & \\
		  &  &  &  & \ltemp &  & \rho_{7,7} & & & & & \rtemp & & & & \\
		  &  &  &  & \ltemp & \rho_{6,8} & & \rho_{8,8} & \rho_{8,9} & & \rho_{8,11}& \rtemp & & & & \\
		  &  &  &  & \ltemp & \rho_{6,9} & &\rho_{8,9} & \rho_{9,9} & & \rho_{9,11} & \rtemp & & & & \\
		  &  &  &  & \ltemp &  & & & & \rho_{10,10} & & \rtemp & & & & \\
		  &  &  &  & \ltemp &  \rho_{6,11} &  & \rho_{8,11} & \rho_{9,11} & & \rho_{11,11} & \rtemp & & & & \\ \cline{6-15} 		    
		  &  &  &  &  &   &  &  & & & \ltemp & \rho_{12,12} & & \rho_{12,14} & & \rtemp \\
		  &  &  &  &  &   &  &  & & & \ltemp &  & \rho_{13,13}& & \rho_{13,15} & \rtemp \\
		  &  &  &  &  &   &  &  & & & \ltemp & \rho_{12,14} & & \rho_{14,14}& & \rtemp \\
		  &  &  &  &  &   &  &  & & & \ltemp &  & \rho_{13,15} &  &  \rho_{15,15}& \rtemp \\ \cline{12-16}
		  &  &  &  &  &   &  &  & & &  &  & & & \ltemp & \rho_{16,16}  \\ 
	\end{array}\right)
    \end{equation}
\end{widetext}
 {Based on the 2s-RDM, any bipartite density matrix-based entanglement measure between two lattice sites $i,\, j$ can be calculated.} Since our calculations are at finite temperatures, we consider entanglement of mixed states.
As there is not one definitive measure for entanglement, we will proceed by focusing on an "upper bound" and a "lower bound" for entanglement, such that both measures combined provide a complete picture.  {Here, we restrict ourself to a quick introduction of the applied entanglement measures, for detailed reviews of the topic we refer the reader to Refs.~\onlinecite{Horodecki_2009,amico_entanglement_2008,Friis_2019_ent}.}

As the first entanglement measure, we consider the mutual information \cite{zurek_information_1983,barnett_entropy_1989}.
The mutual information is defined through the von Neumann entropy as a measure of all correlations in a system
\begin{equation}\label{eq_I}
    I_{ij}\equiv \textrm{tr}\left[ \, \rho_{ij}\ln  \rho_{ij}  \, \right] - 2 \textrm{tr}\left[\, \rho_{i}\ln \rho_{i} \, \right],
\end{equation}
where $\rho_i=\textrm{tr}_j\rho_{ij}=\rho_j=\textrm{tr}_i\rho_{ij}$ is the one-site reduced density matrix.
$I=0$ corresponds to a separable and completely uncorrelated system, while $I>0$ indicates entanglement but also classical correlations between $i$ and $j$. We elaborate on this by showing analytical results for the antiferromagnetic Heisenberg dimer in Appendix~\ref{I_problem}. Since the mutual information does not discriminate between quantum and classical correlations, it is an upper bound to entanglement, in the sense that zero mutual information implies no correlation and consequently no entanglement.
A maximum is reached at $I=2\ln\left(\textrm{dim} \, \rho_i\right)=2\ln\left(4\right)$, with 4 spin and charge degrees of freedom per site. 

For comparison with QMC, the Rényi mutual information \cite{renyi_measures_1961,grover_entanglement_2013} is computed as well
\begin{equation}\label{eq_IR}
    I^R _{ij} \equiv \ln \left[\textrm{tr}  \rho_{ij}^2  \right] - 2 \ln \left[\textrm{tr} \rho_{i}^2  \right].
\end{equation}
The latter differs from the mutual information by the exponent by which $\rho$ contributes,  here 2, i.e.,\   $\rho^2$ \footnote{The Rényi mutual information is based on the generalized Rényi entropy $S_{\alpha}=\ln(\textrm{tr}\rho^{\alpha})/(1-\alpha)$.  Our definition of the Rényi mutual information eq.~\eqref{eq_IR} corresponds to $\alpha = 2$. For $\alpha = 1$ the original von Neumann entropy and its mutual information eq.~\eqref{eq_I} is recovered.}. Due to its possible negativity \cite{Kormos_2017,kudler-flam_renyi_2023}, $I^R$ is not a well-defined entanglement measure.

Besides the "upper bound" $I_{ij}$, we consider as a "lower bound" the negativity $N$.
It investigates the physicality of the density matrix under the application of the partial transpose on site $j$ only. Here, the partial transpose acts on any operator $O$ in the Hilbertspace of $i\cup j$, and is elementwise defined as
\begin{equation}\label{partial_transpose}
    \langle \psi_i, \psi_j | O^{T_j} | \bar{\psi}_i, \bar{\psi}_j \rangle \equiv  \langle \psi_i,  \bar{\psi}_j |O | \bar{\psi}_i, \psi_j \rangle.
\end{equation}
 {Here, $\psi_i,\bar{\psi}_i,\psi_j,\bar{\psi}_j,$ are four (possibly different) state vectors on sites $i,j$. For the partial transpose of $\rho$, this transformation is performed on all matrix elements, i.e., all physically allowed combinations of basis vectors $v$ defined in Table~\ref{tab:check}. }
If a density matrix is separable $\rho_{ij}=\rho_i\otimes\rho_j$, its partial transpose $\rho_{ij}^{T_j}=\rho_i\otimes\rho_j^{T}$ remains a physical density matrix and is positive semi-definite. Conversely, if the partial transpose $\rho_{ij}^{T_j}$ is not physical, the density matrix is not separable \cite{peres_separability_1996,horodecki_separability_1996}.
To detect an unphysical partially transposed density matrix, one investigates if $\rho_{ij}^{T_j}$ is positive semi-definite, by summing over all its negative eigenvalues $\lambda$ 
\begin{equation}\label{eq_negativity}
    N \equiv - \sum_{\lambda<0} \lambda.
\end{equation}
If the negativity $N$ is non-zero, the original matrix $\rho_{ij}$ is not separable {, this is known as Peres-Horodecki or positive partial transpose criterion. }
The opposite statement is not true, since bound entangled states may exist with $N=0$. Therefore, the negativity provides a lower bound to entanglement  {in the sense that a non-zero negativity proves the existence of entanglement \cite{footnoteN}}.
Oftentimes, the negativity is rewritten in terms of the trace norm  {$\| \rho \|_{\textrm{tr}} = \textrm{tr}\sqrt{(\rho \rho^{\dagger})}$ }as  \cite{vidal_computable_2002}
\begin{equation}\label{eq_tr_N}
    N = \frac{\| \rho_{ij}^{T_j} \|_{\textrm{tr}}-1}{2}.
\end{equation}
The measure is closely related to the logarithmic negativity $E_N = \log_2\| \rho_{ij}^{T_j} \|_{\textrm{tr}}$ \cite{plenio_logarithmic_2005,vidal_computable_2002}.
Unfortunately, the negativity cannot be directly applied to fermionic systems  \cite{eisert_entanglement_2018}, as illustrated in Appendix~\ref{Measure_Comparison}. 

Instead of the negativity, we thus consider a recent extension thereof, based on the partial time-reversal transformation \cite{shapourian_partial_2017,shapourian_entanglement_2019}.
We adapt the definition of the partial time-reversal operation from Shapourian \textit{et al.} \cite{shapourian_partial_2017} to the case of a two-site density matrix with spin $1/2$ particles.
The partial time-reversal operation applied on an operator $O$ yields
\begin{equation}
    \langle \psi_i, \psi_j | O^{{\rm TR}_j} | \bar{\psi}_i, \bar{\psi}_j \rangle \equiv  \langle \psi_i,  \bar{\psi}_j |O | \bar{\psi}_i, \psi_j \rangle (-1)^{\phi}
\end{equation}
where the phase factor 
\begin{equation}
\begin{aligned}
\phi = & \frac{n_i\left(n_i+2\right)}{2}+\frac{\bar{n}_i\left(\bar{n}_i+2\right)}{2}+\bar{n}_j n_j \\
& +n_i n_j+\bar{n}_i \bar{n}_j+\left(\bar{n}_i+\bar{n}_j\right)\left(n_i+n_j\right)
\end{aligned}
\end{equation}
is given in terms of the number of electrons $n_i = n_{i,\uparrow} + n_{i,\downarrow}$ of each matrix element. For example $|\uparrow,\uparrow\downarrow\rangle\langle\uparrow,\uparrow\downarrow|$ has $n_i=\bar{n}_i=1$ and $n_j=\bar{n}_j=2$. 
In simple terms, this accounts for the number of minus signs picked up by permuting the fermionic operators in the process of taking the partial transpose.
Equivalently to Eq.~\eqref{eq_tr_N}, the fermionic negativity is
\begin{equation}
    N^F \equiv \frac{\| \rho_{ij}^{\rm{TR_j}} \|_{\textrm{tr}}-1}{2}.
\end{equation}
And similar to the partial transpose,  {$N^F$ is only non-zero if $\rho_{ij}$ is not separable.} Therefore, a non-zero fermionic negativity detects an entangled state. Moreover, one expects that any entangled state will necessarily lead to a nonzero fermionic negativity \cite{shapourian_entanglement_2019}. 

In Appendix~\ref{Measure_Comparison}, we compare results from the four measures discussed here, i.e., $I$, $I^R$, $N$, and $N^F$, as well as the non-freeness \cite{gottlieb_properties_2006,held_physics_2013,bellomia2025localclassicalcorrelations}.

\section{Model and Method} \label{Model_and_Method}
We investigate the one-band Hubbard model with nearest neighbor (NN) hopping $t$
\begin{equation}
\begin{split}
    H =& -\sum_{\langle ij\rangle,\sigma} t \left( c^{\dagger}_{i,\sigma}c_{j,\sigma} + c^{\dagger}_{j,\sigma}c_{i,\sigma} \right) \\ 
    & + U\sum_{i} n_{i,\uparrow}n_{i,\downarrow} -\mu \sum_{i,\sigma} n_{i,\sigma},
\end{split}
\end{equation}
where $\mu$ is the chemical potential, $U$ the onsite Coulomb repulsion, $n_{i,\sigma} = c_{i,\sigma}^{\dagger}c_{i,\sigma}$ the occupation number operator, and $\hat{c}_{i\sigma}^{(\dagger)}$ is the fermionic annihilation (creation) operator that annihilates (creates) an electron at site $i$ with spin $\sigma$. We consider the square lattice in the paramagnetic phase, which implies $SU(2)$ symmetry.  {For a more detailed introduction of the Hubbard model see for example Refs.~\onlinecite{qin_hubbard_2022,HalTasaki_1998, SchaeferPRX}.} The Fourier transform of the kinetic term gives, for the NN hopping, the tight-binding dispersion relation
\begin{equation}\label{eq_tight_binding_epsilon}
    \varepsilon_{\mathbf{k}}=-2t(\cos k_x + \cos k_y),
\end{equation}
where the momentum ${\bf k} = (k_x, k_y)$ is measured in units of the inverse of the lattice constant $a$. In the following, we set $\hbar = k_B=a=1$ and measure all quantities in untis of $t\equiv 1$.

To solve the Hubbard model, $p$D\textGamma A in the multi boson exchange formalism is applied \cite{krien_tiling_2021}. For vertices and the self-energy, calculations are performed on a $16\times 16$ momentum grid with periodic boundary conditions. A 10 times finer momentum grid is applied for sums over one-particle Green's function (coarse graining~\cite{iskakov_single-_2022}). For comparison with QMC, data without coarse graining is used.
The difference in the entanglement measures between results with and without coarse graining is shown in Appendix~\ref{coarse_graining}.

\subsection{Parquet dynamical vertex approximation}

The $p$D\textGamma A  method uses as its basis the local two-particle fully irreducible vertex $\Lambda$, which is obtained from the  solution of the DMFT impurity problem.  In this work, the DMFT was solved with the continuous-time QMC in the hybridization expansion as implemented in the {\it w2dynamics} code~\cite{w2dyn}. The resulting $\Lambda$ is then used as input to the parquet equations. 

The parquet equations are a set of exact relations between different classes of two-particle vertices and between the self-energy and the full two-particle vertex~\cite{Dominicis64, Dominicis64-2}. A good introduction to the formalism is provided in~\cite{Bickers04}. In this work, we use data generated with the multi-boson exchange implementation of the method as presented in~\cite{krien_tiling_2021,krien_plain_2022}. This allows for a careful treatment of the Matsubara frequency asymptotics needed in calculations of infinite sums. For some of the values of the parameters (i.e., for $U=2$ and $U=3$ out of half-filling), the parquet approximation (PA) was used. In the PA, the fully irreducible vertex $\Lambda$ is approximated as $U$. While for weak coupling at $U=2$, the $p$D\textGamma A is essentially reduced to PA for the range of the temperatures considered ($\Lambda$ from DMFT is practically equal to $U$), for $U=3$ this is not the case and we show results for both, $p$D\textGamma A and PA, in the half-filled case.

The $p$D\textGamma A is the computationally most involved diagrammatic extension of the DMFT~\cite{rohringer_diagrammatic_2018} and therefore currently only relatively high temperatures and small grid sizes are available (there are, however, promising computational developments~\cite{Rohshap2025}). The reason for this is that the vertices retain their full frequency and momentum dependence that reflects best spatial correlations. The method treats, by construction, all scattering channels equally, as well as the interplay between them. In cases where a clear dominance of one scattering channel is present, a simpler approach that retains dependence on one momentum only -- the transfer momentum in the dominant channel -- can be used. This simplification lies at the core of the ladder D\textGamma A method~\cite{rohringer_diagrammatic_2018}.

\subsection{Calculating 2s-RDM from $p$D\textGamma A output}\label{quantities}

The 2s-RDM can be calculated as a postprocessing step using the method developed in Ref.~\onlinecite{roosz_two-site_2024}, where the 
Heisenberg equation of motion links all symmetry-allowed matrix elements of $\rho_{ij}$ to frequency sums over two- and four-point Green's functions. While all the details are provided in Ref.~\onlinecite{roosz_two-site_2024} and the relation between Green's functions and 2s-RDM recapped in Appendix~\ref{2srdm_from_G_F}, we briefly sketch here which quantities obtained from $p$D\textGamma A are used in calculations in this work.

For completeness and also to introduce the notation, we start with the definition of the one-particle Green's function
\begin{equation}
    G^{k} = - \sum_{i,j} \int_0^{\beta} d\tau\ e^{i\nu\tau}e^{-i\mathbf{k}(\mathbf{r}_i-\mathbf{r}_j)} \left\langle T_\tau \hat{c}_{i,\sigma}(\tau) \hat{c}_{j,\sigma}^{\dagger}(0)\right\rangle,
\end{equation}
with $\tau$ denoting the imaginary time and $\beta\equiv 1/T$ the inverse temperature.  We use a joint four-vector notation in which $k = (\nu,\mathbf{k})$ labels the combination of discrete fermionic Matsubara frequency $\nu_n = \frac{(2n+1)\pi}{\beta}$, $n \in \mathbb{Z}$, and momentum vector $\mathbf{k}$. We omit the spin index of $G^{k}$, since due to $SU(2)$ symmetry the one-particle Green's function is the same for both spins.

Analogously, we define the two-particle (four-point) Green's function
\begin{equation}
    \begin{aligned}
G&_{\sigma_1 \ldots \sigma_4}^{kk'q}  = \sum_{i,j,l,m} e^{i \mathbf{k} \mathbf{r}_i} e^{-i(\mathbf{k}+\mathbf{q}) \mathbf{r}_j} e^{i(\mathbf{k}^{\prime}+\mathbf{q}) \mathbf{r}_l} e^{-i \mathbf{k}^{\prime} \mathbf{r}_m}   \\
& \times \int_0^\beta d \tau_1 \int_0^\beta d \tau_2 \int_0^\beta d \tau_3 e^{i \nu \tau_1} e^{-i(\nu+\omega) \tau_2} e^{i(\nu^{\prime}+\omega) \tau_3} \\
& \times\left\langle T_\tau \hat{c}_{i,\sigma_1}\left(\tau_1\right) \hat{c}_{j,\sigma_2}^{\dagger}\left(\tau_2\right) \hat{c}_{l,\sigma_3}\left(\tau_3\right) \hat{c}_{m,\sigma_4}^{\dagger}(0)\right\rangle  ,
\end{aligned}
\end{equation}
where we additionally have a bosonic frequency-momentum index  $q = (\omega,\mathbf{q})$, with bosonic Matsubara frequency  $\omega_n= \frac{2\pi n }{\beta}$ and momentum ${\bf q}$. Due to $\mathrm{SU}(2)$ symmetry, the number of spin components that need to be computed reduces to the following:  $G_{\sigma\sigma\sigma'\sigma'}$, which we will denote by $G_{\sigma\sigma'}$, and $G_{\sigma(-\sigma)(-\sigma)\sigma}$, which can be shown to be equal to $G_{\sigma\sigma} - G_{\sigma(-\sigma)}$. Furthermore, since $G_{\sigma\sigma'} = G_{(-\sigma)(-\sigma')}$,
we only need to compute $G_{\uparrow\uparrow}$ and $G_{\uparrow\downarrow}$~~\cite{rohringer_diagrammatic_2018}.

 {To compute all elements of the 2s-RDM it is enough to know the one- and two-particle Green's functions. These quantities can be reconstructed from the output of $p$D\textGamma A. It is however more convenient and computationally cheaper to use the direct output of $p$D\textGamma A in the calculations, because many of the expressions for the correlators needed in $\rho_{i,j}$ contain summation over momenta or frequencies, which reduce the two-particle Green's functions to smaller objects (dependent on fewer arguments), like susceptibilities $\chi$, screened interaction $W$ and the fermion-boson vertex $\gamma$. Only some elements require the knowledge of the full two-particle vertex $F$, which has the same number of arguments as the two-particle Green's function. Below, we define all the quantities that were taken from the $p$D\textGamma A calculations and used in expressions in Appendix~\ref{2srdm_from_G_F} to obtain the 2s-RDM.}

The full two-particle vertex $F$ is defined by subtracting disconnected parts of the four-point Green's function and by "amputating the legs", which results in the following expression \cite{rohringer_diagrammatic_2018}
\begin{align}
    {G}_{\sigma\sigma'}^{kk'q} &= G^k G^{k'}\delta_{q0} - G^k G^{k+q}\delta_{kk'}\delta_{\sigma\sigma'} \nonumber \\
    &- G^k G^{k+q}{F}_{\sigma\sigma'}^{kk'q}G^{k'}G^{k'+q}.
\end{align}
To make the connection between the vertex and the physical susceptiblities, we introduce the following spin combinations: magnetic $X_m = X_{\uparrow\uparrow} - X_{\uparrow\downarrow}$ and density $X_d = X_{\uparrow\uparrow} + X_{\uparrow\downarrow}$ where $X$ is any two-particle quantity such as $G$, $F$ or the susceptibility $\chi$. 
That is, in this notation, the physical susceptibility in the $\alpha=m,d$ channel is given by
\begin{equation}\label{eq_chi}
    \chi_\alpha^q=-2 \sum_k G^k G^{k+q}-2 \sum_{k, k^{\prime}} G^k G^{k+q} F_\alpha^{k k^{\prime} q} G^{k^{\prime}} G^{k^{\prime}+q}.
\end{equation}
Following Ref.~\onlinecite{roosz_two-site_2024}, we can directly compute some contributions to the 2s-RDM from the susceptibilities. Other contributions require
 the screened interaction $W^q_\alpha$ defined as \cite{krien_single-boson_2019,krien_tiling_2021}
\begin{equation}
    W^q_\alpha = U_\alpha -  \frac{1}{2}U_\alpha\chi^q_\alpha U_\alpha,
    \label{eq:Wdef}
\end{equation}
where $U_d = U$, $U_m = -U$. The susceptibility $\chi$ and $F$ are connected to the fermion-boson irreducible vertex $\gamma_\alpha^{k q}$ through~\cite{krien_single-boson_2019}
\begin{align}
  &\gamma^{kq}_\alpha = \frac{1 + \sum_{k'}
   F_\alpha^{kk'q}G^{k'}G^{k'+q} }{1-\frac{1}{2}U_\alpha\chi_\alpha^q}.
   \label{eq:gamma_def}
\end{align}
Both $W$ and $\gamma$ are direct output of the  $p$D\textGamma A in the multi-boson exchange implementation~\cite{krien_tiling_2021,krien_plain_2022}. 
 {As mentioned above, in principle, all quantities required for the 2s-RDM can be expressed only through $F$ and $G$, as can be seen from the above equations. Using the susceptibility $\chi$, screened interaction $W$ and Fermi boson vertex $\gamma$ has the advantage that their memory requirement scales with the number of momenta $N_{k_{x/y}}$ and fermionic frequencies $N_F$ as $\mathcal{O}(N_{k_x}N_{k_y}N_F)$, $\mathcal{O}(N_{k_x}N_{k_y}N_F)$, $\mathcal{O}(N_{k_x}^2N_{k_y}^2N_F^2)$ respectively, compared to the scaling $\mathcal{O}(N_{k_x}^3N_{k_y}^3N_F^3)$ of $F$. Hence, $p$D\textGamma A is most efficiently performed in the single boson exchange formalism since the parquet equations can be formulated in terms of these quantities \cite{krien_tiling_2021}.  
Further noting that $F$ at high frequencies can be expressed through $\gamma$ and $W$, we exploit the difference in scaling and replace $F$ by:
\begin{align}
    F_{\alpha}^{kk'q} = T_{\alpha}^{kk'q} + \gamma_\alpha^{kq}W_\alpha^q \gamma_\alpha^{k'q},
    \label{eq:Tgwg}
\end{align}
where the memory-intensive multi-boson exchange term $T$ is only computed on small frequency boxes. The memory requirement of this $T$ is also the limiting factor for the computation of the 2s-RDM, however the bottleneck for the formalism is the complexity of solving the parquet equations in $p$D\textGamma A. As a rule of thumb, the number of required frequencies scales as $N_F \sim  \beta$ and the number of momenta scales with the correlation length $\xi$ as $N_{k_x}=N_{k_y}\sim e^{\xi}$. The decomposition of $F$ into $T+\gamma W \gamma$ } is particularly important for contributions to the 2s-RDM coming from derivatives of the four-point Green's function that result in the full vertex multiplied by frequencies -- the additional two-point Green's functions that appear in those contributions guarantee that these terms do not diverge, but a careful asymptotic treatment as explained in the Appendix of Ref.~\onlinecite{roosz_two-site_2024} is required. 

\begin{figure*}[h!t]
\includegraphics[width=\linewidth]{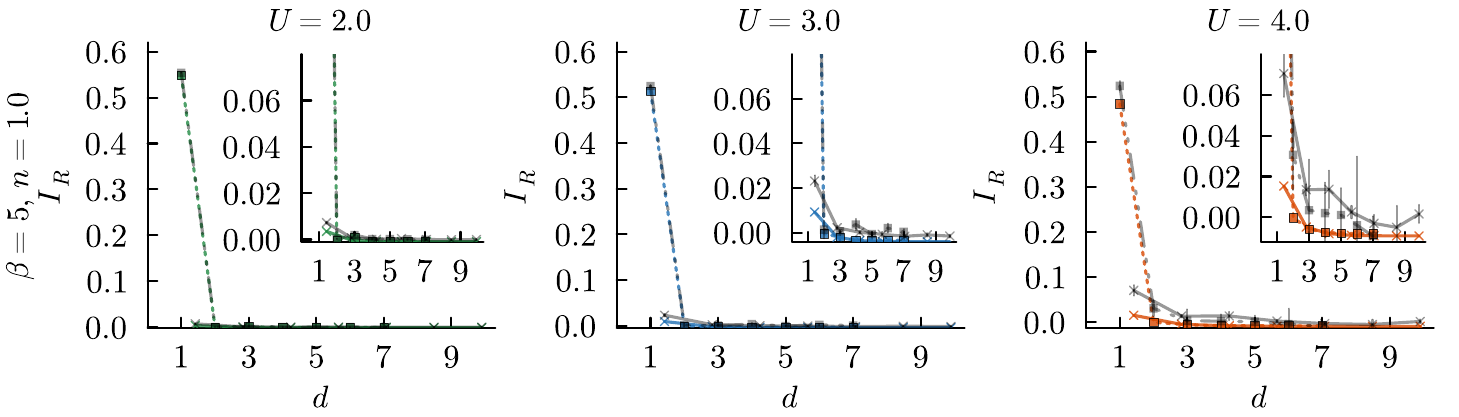}
\caption{Rényi mutual information $I_{\textrm{R}}$ as
a function of distance $d$ comparing $p$D\textGamma A (color) and QMC (grey)  
for the 16$\times$16-site Hubbard model at half-filling and, from left to right,  $U=2,\ 3,\ 4$. Solid lines and crosses are measurements along the diagonal of the lattice $\mathbf{\Delta}=(\Delta,\Delta)$, boxes and dotted lines are measured along the NN path $\mathbf{\Delta}=(\Delta,0)$. We denote the real distance $d=\sqrt{\Delta_x^2+\Delta_y^2}$ from the reference site on the $x$-axis. Insets are the same plot with an enlarged y-axis. For $U=2,3$ the QMC error bars are smaller than the symbols. Errors only become significant in the inset for $U=4$, here both the statistic and Trotter error are relevant.
}
\label{Fig_QMC_Comparison}
\end{figure*}

\begin{figure*}[h! t]
\includegraphics[width=0.495\linewidth]{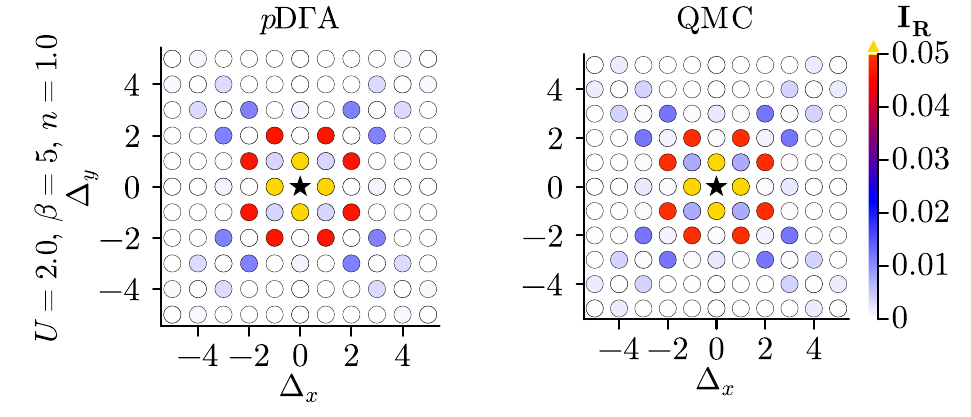}
\includegraphics[width=0.495\linewidth]{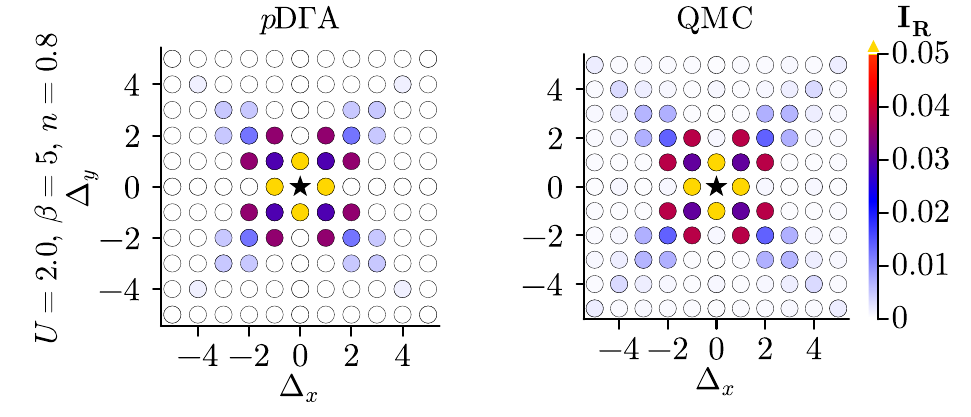}
\caption{Rényi mutual information $I_{\textrm{R}}$ (false colors) between a reference point (star) and a
second lattice site at distance $(\Delta_x,\Delta_y)$, comparing $p$D\textGamma A  and QMC for $U=2$, $\beta=5$ and two fillings, i.e., $n=1$ (half-filling, two leftmost panels) and $n=0.8$ (two rightmost panels)~\cite{footnoteRS}. 
Note that NN $I^R$ is outside the color bar and marked in yellow;
from left to right we have  for NN: $I^{R}_{01} = 0.5484$, $I^{R}_{01}=0.5540$; $I^{R}_{R,01} = 0.4982$, $I^{R}_{R,01}=0.5031$.}
\label{Fig_QMC_Comparison_2D}
\end{figure*}


\section{Comparison to quantum Monte Carlo}\label{QMC_comparison_main}
In this section, we benchmark the results for the mutual information based on Rényi entropy obtained from 2s-RDM with $p$D\textGamma A  against direct Quantum Monte Carlo calculation. Since $p$D\textGamma A is not an exact method, there will be some deviations to QMC, especially at larger $U$ and lower $T$.
 For $\beta=5$ and $U=2$, on the other hand,  it has been shown that there is quantitative agreement between the parquet equations result (even in parquet approximation) and QMC~\cite{Hille2020, SchaeferPRX}, whereas for $U=3$ differences start to be visible~\cite{Hille2020}. Here we use the same parameters and compare the Renyi mutual information for a $16\times16$ lattice with periodic boundary conditions.  

\subsection{Obtaining Rényi entropy from QMC}

QMC methods solve the many-body problem by stochastically sampling the high-dimensional integral of the grand canonical ensemble. Where sign-problem-free formulations exist, this approach allows for error-controlled solutions, making it the method of choice for benchmarking other numerical methods.
We use the algorithms for lattice fermions (ALF) package, which provides an implementation of the finite-temperature and projective auxiliary-field quantum Monte Carlo algorithm~\cite{assaad_alf_2022}, which constructs the Rényi mutual information from two independent simulations of the system~\cite{grover_entanglement_2013,Assaad13a}.
For the calculations we have used  a symmetric  Trotter decomposition with an  imaginary time step set  to
$\Delta\tau t  = 0.1$. At $U/t=4$  Fig. 2 of Ref.~\onlinecite{assaad_alf_2022}   shows that this produces a negligible systematic error for the energy. 

\subsection{Comparison between $p$D\textGamma A and QMC}

To compare the two numerical methods, we compute the Rényi mutual information between two lattice sites separated by $\mathbf{\Delta}\equiv\mathbf{r}_j-\mathbf{r}_i = (\Delta_x,\Delta_y)$. In the entire comparison, we do not apply coarse graining in $p$D\textGamma A as QMC simulates a finite $16\times16$ system.

Fig.~\ref{Fig_QMC_Comparison} shows the Rényi mutual information $I_R$ at half-filling ($n_i=n=1$) and $\beta = 5$, for different values of $U$ as a function of distance $d=\sqrt{\Delta_x^2+\Delta_y^2}$. Dashed lines show $I_R$ along the $x$-direction $\mathbf{\Delta}=(\Delta,0)$ and solid lines are data along the diagonal  $\mathbf{\Delta}=(\Delta,\Delta)$.

While the  Rényi mutual information is large between NN neighbors, it drops off extremely rapidly 
already for the next-next nearest neighbor, necessitating the enlargement of the insets.
This suggests only a weak entanglement, except for neighboring sites.
Interestingly, we encounter a rarely observed negative Rényi mutual information at large distances.

As for the comparison between $p$D\textGamma A and QMC, the agreement for the smallest interaction $U=2$ is very good, validating the developed $p$D\textGamma A calculation of the  Rényi mutual information.
For $U=4$ and to a lesser extent $U=3$, we find some quantitative though not qualitative discrepancy.
The reasons for the observed differences are: 
(i) The computation of the 2s-RDM  from $p$D\textGamma A requires infinite Matsubara sums and thus the knowledge of the two-particle vertex in a sufficiently large frequency range. For higher values of $U$ this range increases and the finite-box error grows. 
This deficiency can hopefully be cured with the application of novel vertex compression methods~\cite{Rohshap2025}. 
(ii) With larger $U$ 
there will be  corrections to the the $p$D\textGamma A  approximation of a local fully irreducible vertex. This is actually more severe for a small $16\times16$ cluster, than for an infinite lattice, since the momentum dependence of the fully irreducible vertex is more pronounced in small systems~\cite{Wieser2025}.
 
(iii) For $U=4$, $\beta=5$ the antiferromagnetic susceptibility is already very strongly enhanced to $\chi(\omega_n=0,\mathbf{q}=({\pi,\pi})) \approx 30$ in QMC. In DMFT, antiferromagentic order would have already set in (not shown).  In the temperature regime where the susceptibility strongly increases, only a minor deviation of the onset temperature for this increase will result in largely different susceptibilities.
Such a difference is observed for the susceptibility in Appendix~\ref{comp_chi} for $U=4$; and Fig.~\ref{Fig_QMC_Comparison} shows how
this reflects in the  Rényi mutual information.


In Fig.~\ref{Fig_QMC_Comparison_2D}, we also present a real space map of the results for $U=2$ at two different fillings $n=1.0$ and $0.8$. The agreement between
$p$D\textGamma A and QMC is again very good, but for further neighbors the 
$p$D\textGamma A mutual information is slightly smaller, particularly upon doping.   Note that the filling $n=0.8$ is also observed at slightly different values of the chemical potential $\mu_{\textrm{QMC}}=0.2610$, $\mu_{\textrm{D\textGamma A}}=0.2667$.
Interestingly, at half-filling there is an alternating pattern of sites with more and less  Rényi mutual information, and thus supposedly entanglement. All of these alternations are however small compared to the NN $I^R$ which has by far (even on another scale) the largest mutual information. This also holds for the conventional (non-Rényi)  mutual information and other entanglement measures, and will be discussed in more detail in the next Section.


\begin{figure*}
\includegraphics[width=\linewidth]{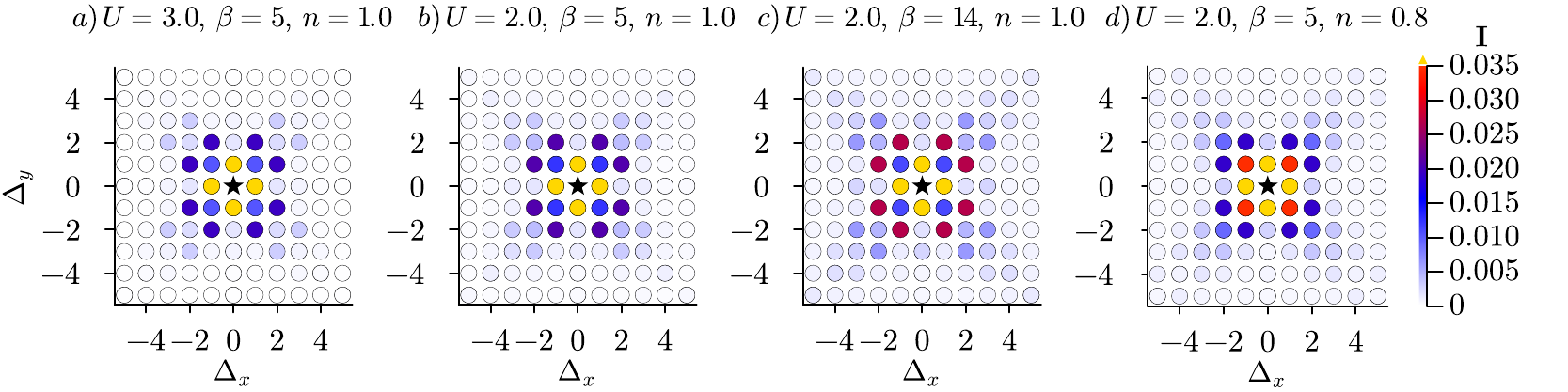}
\par\vspace{0.35cm}
\includegraphics[width=\linewidth]{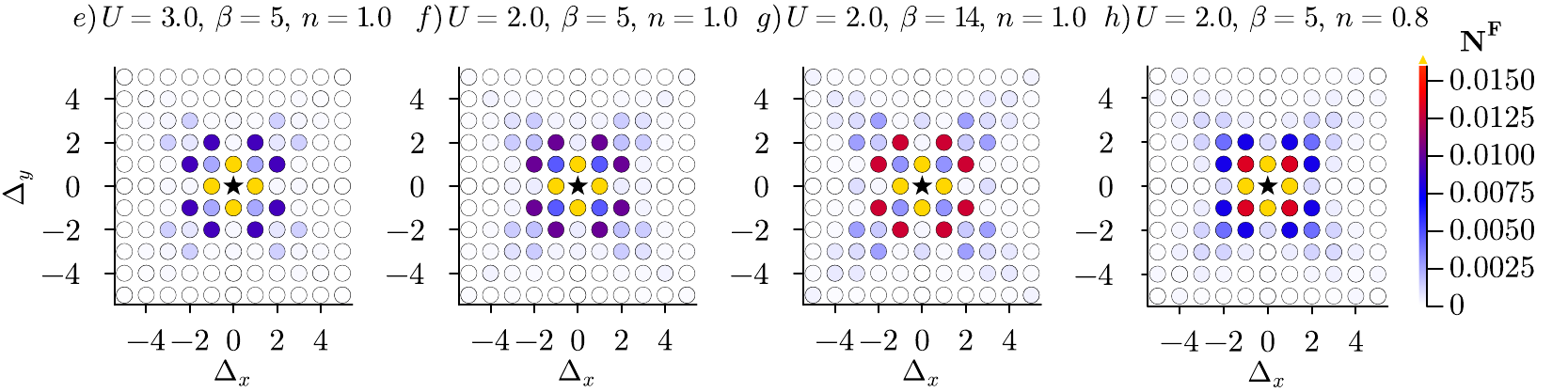}
\caption{a)-d) Mutual information $I$ and e)-h) fermionic negativity $N^F$ plotted as real space entanglement between lattice sites separated by $\mathbf{\Delta}=(\Delta_x,\Delta_y)$ from a reference site marked by $\star$. From left to right we display (a,e) $U=3$, $\beta=5$, $n=1$ where the yellow NN values are $I_{01} = 0.3014$, $N^F_{01}=0.1646$; (b,f) $U=2$  $\beta=5$, $n=1$ with $I_{01} = 0.3114$, $N^F_{01}=0.1663$; (c,g) $U=2$, $\beta=14$, $n=1.0$ with $I_{01} = 0.3196$, $N^F_{01}=0.1732$; (d,h) and $U=2$, $\beta=5$, $n=0.8$ with $I_{01} = 0.3181$, $N^F_{01}=0.1417$ \cite{footnoteRS}.}
\label{Fig_structure}
\end{figure*}

\begin{figure}
\includegraphics[scale=1]{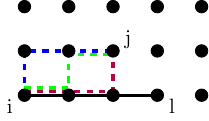}
\caption{Three different paths (dashed colored lines) connect sites $i$ and $j$ with three hoppings each, whereas there is just one path that connects $i$ and $l$ (solid line) with three hoppings. The Manhattan distance between $i$ and $j$ is the same as that between $i$ and $l$, the Euclidean distance is even larger for the former pair.}
\label{Fig_manhattan}
\end{figure}

\begin{figure*}
\includegraphics[width=\linewidth]{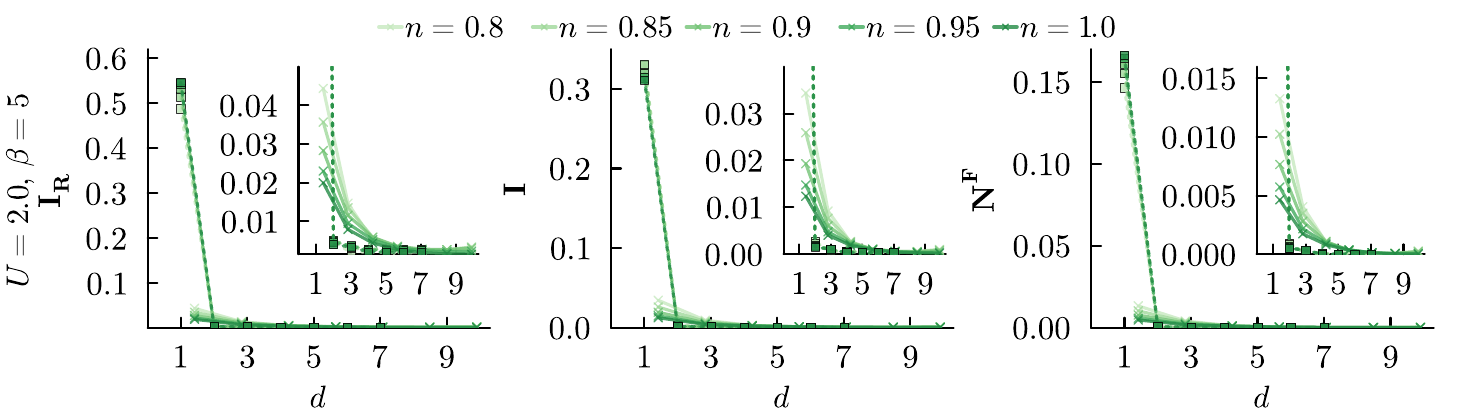}
\includegraphics[width=\linewidth]{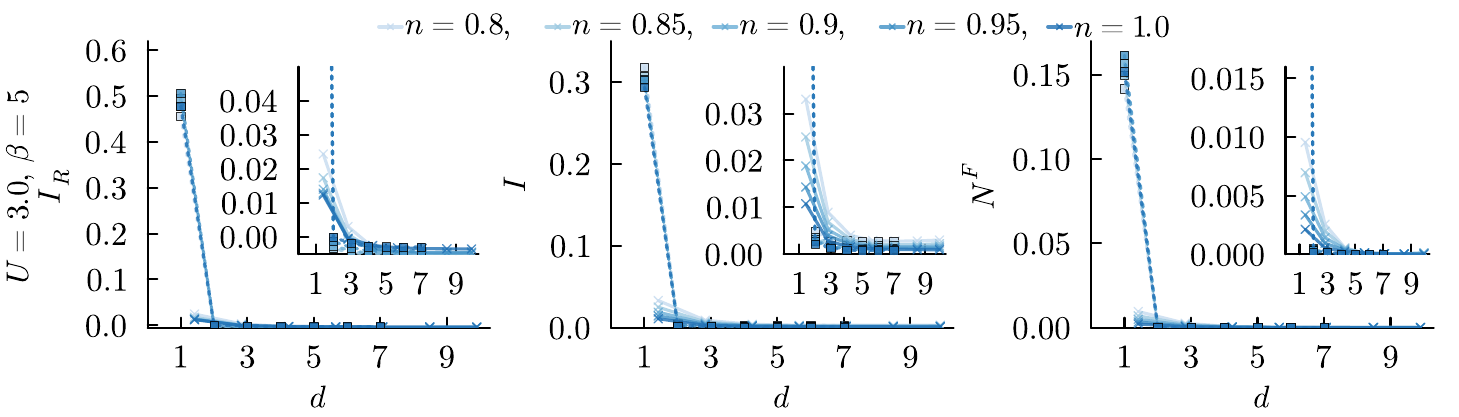}
\caption{Comparison of the 
Rényi mutual information $I_R$ (left),  mutual information $I$ (middle), and fermionic negativity $N^F$ (right)
at disttance $d$ for different fillings $n$; $U=2$ (top) and $U=3$ (botttom) at $\beta=5$.
From left to right we show Rényi mutual information $I_R$, mutual information $I$, and fermionic negativity $N^F$.
Solid lines and crosses are measurements along the diagonal of the lattice while boxes and dotted lines are measured along the $x$ (or $y$) direction. Insets are the same plot with an enlarged y-axis.
}
\label{Fig_n_Comparison}
\end{figure*}


\section{Results}\label{Results}
Having tested and demonstrated the reliability of 2s-RDM results obtained from $p$D\textGamma A in the previous section, we now turn to an analysis of the bi-partite  {correlation and} entanglement structure of the 2D Hubbard model. We restrict ourselves to nearest-neighbor (NN) hopping on the square lattice in the weakly correlated Fermi liquid phase.

In Fig.~\ref{Fig_structure}, we investigate the structure of entanglement within real space for different values of interaction strength $U$, filling $n$, and inverse temperatures $\beta$. 
To obtain meaningful results, we combine the von Neumann mutual information ($I$, instead of the Rényi mutual information of the previous section; top) as an upper bound to entanglement, and the quantitatively unrelated fermionic negativity ($N^F$; bottom) as a lower bound.
Both measures follow the same qualitative behavior, proving that the observed structure indeed resembles the spatial structure of entanglement.

Entanglement is strong only for NN sites and rapidly decreases by approximately one order of magnitude for larger distances. However, this decrease is neither isotropic nor smooth as one might na\"ively expect or has in a free electron gas scenario. In fact, entanglement is modulated by two main effects.    

First, as a universal feature of our results, we observe that entanglement along the NN direction $\mathbf{\Delta}=(\Delta,0)$ is less long-ranged than entanglement close to the diagonal direction $\mathbf{\Delta}=(\Delta,\Delta)$. This is a direct consequence of the NN hopping $t$ and the connectivity between different sites. 
The connectivity of two sites is given by the number of paths that connect them with a specific number of NN hoppings. Comparing two sites that are separated from a reference site by the same number of NN hoppings (i.e., same "Manhattan distance"), we find that the connectivity grows the more diagonal a path becomes. This is illustrated for two example sites in Fig.~\ref{Fig_manhattan}.
Hence, sites that are connected by diagonal paths are expected to be more strongly entangled.

Furthermore, we observe a second attenuating effect. Even in the weak-coupling Fermi liquid phase ($U=2,3,4$), far from the antiferromagnetic phase and Mott transition, entanglement at half-filling exhibits a checkerboard pattern, reflecting the inherent antiferromagnetic tendency of the square lattice NN-hopping Hubbard model and resembling Néel order itself.
Entanglement favors sites whose spins have an antiferromagnetic alignment (compare the real space spin susceptibility in Fig.~\ref{fig_susc} with Fig.~\ref{Fig_structure}). 
The same pattern of entanglement remains in the noninteracting case $U=0$ (Appendix~\ref{Mutualinformation_U0}). That is, the  Néel-like pattern
does not necessitate strong interactions, but can also be caused by the perfect nesting of the Fermi surface \cite{ehlers_entanglement_2015} $\varepsilon_{\mathbf{k}+(\pi,\pi)} = -\varepsilon_{\mathbf{k}}$, which leads to a divergent spin susceptibility at $\mathbf{q}=(\pi,\pi)$ \cite{lin_two-dimensional_1987}.

Moreover, when a doping lifts this nesting effect and antiferromagnetism is suppressed,
we observe at both $U=0$ (Fig.~\ref{Fig_non_int_half_fil_muti}) and finite $U$ (Fig.~\ref{Fig_structure} d/h)), that the checkerboard pattern vanishes.
Note that at low filling $n \simeq 0.8$ our method encounters nonphysical density matrices due to finite numerical precision. Where such nonphysical negative eigenvalues were encountered, we optimize for the closest positive semi-definite density matrix with unity trace following Ref.~\onlinecite{smolin_efficient_2012}.
Conversely, larger interactions $U$ act favorably towards the antiferromagnetism, and the checkerboard pattern becomes more pronounced (compare Fig.~\ref{Fig_structure} a/e) vs. b/f)). 
In accordance with the growing correlation length at lower temperatures, we also observe that the alternating entanglement pattern becomes more pronounced and entanglement more long-ranged at $\beta=14$ (compare Fig.~\ref{Fig_structure} b/f) vs. c/g)).

In Fig.~\ref{Fig_n_Comparison}, we turn from the qualitative patterning to the quantitative changes of entanglement. Going away from half-filling 
which suppressed the alternating entanglement pattern in Fig.~\ref{Fig_structure},
we observe in Fig.~\ref{Fig_structure} that the magnitude of entanglement actually increases in all three entanglement witnesses studied.
This is a bit counterintuitive,  at least for the mutual information, since spin correlations are strongest at half-filling. However, the 2s-RDM also includes the charge degree of freedom, which opens an additional channel for entanglement away from half-filling.

\section{Correlation diagnostics}\label{Section_COR_DIAG}
\begin{figure}
\includegraphics[width=\linewidth]{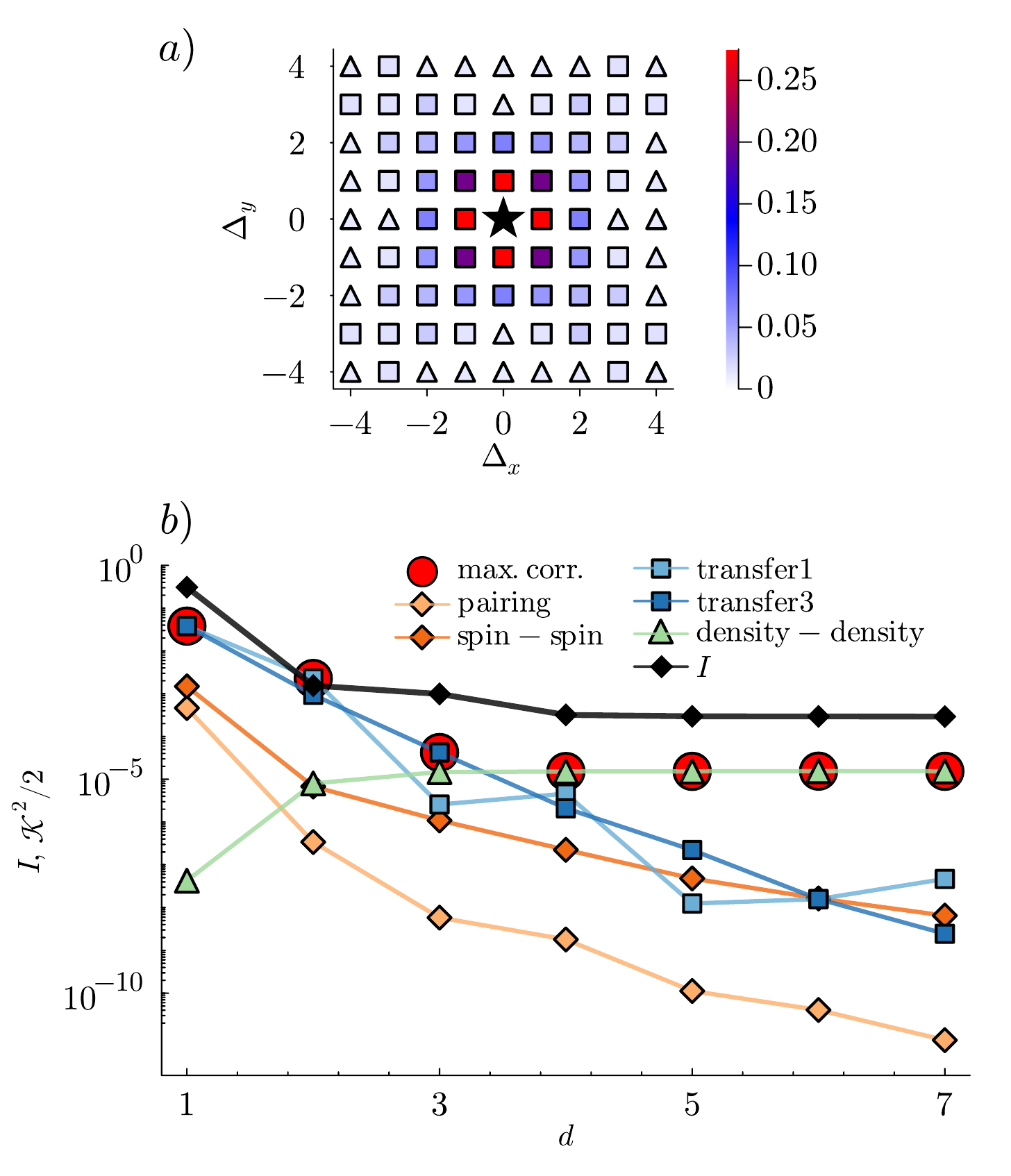}
\caption{Correlation functions for $\beta=5$, $U=2$, half filling. a) Color map of the maximal correlation functions. Squares denote transfer-type correlation, triangles denote density-density correlations. b) ${\cal K}^2/2$ correlators over distance along the NN direction.}
\label{Fig:max_corr_dir}
\end{figure}
 {To better interpret the entanglement results shown in the previous section, we investigate the nature of correlations in the system. Noting that the mutual information is an upper-bound for all correlations $\cal{K}$ in a system \cite{Wolf_2008_MI}, we can interpret the prevalent correlation function in the system to be the dominant contribution to the entanglement which is detected by the mutual information. 
Specifically, knowledge of the reduced density matrix of the two sites allows for the calculation of any expectation value defined on the two sites and any correlation function. Then, information on the nature of the correlation functions as a function of distance, can be obtained by identifying the largest correlation function according to the Frobenius norm \cite{Chincholi_2025} .}

 {Let $O_{(i)}$ be an operator defined at site $i$ and $O_{(j)}$  an operator defined at site $j$, both operators are normalized such that $\textnormal{Tr} O^2_{(i)} = \textnormal{Tr} O^2_{(j)}=1$. We look for the normalized operators that maximize the correlation function
\begin{equation}
    {\cal K} = \langle  O_{(i)} O_{(j)} \rangle - \langle O_{(i)} \rangle \langle O_{(j)} \rangle.
\label{eq:correlation_function}
\end{equation}
 This maximum can be obtained for any two subsystems of any physical system by choosing a basis in the operator space, writing up a matrix for the correlation functions, which are bilinear in the operators, and finding the maximal singular value and the corresponding eigenvalues of the matrix.  The details of the procedure for the general case are given in Appendix \ref{sec:max_corr_function}. 
In a general quantum state, the operators maximizing the correlation function can be any linear combination of all possible operators from the subsystem.
However, in the thermal equilibrium state of the Hubbard model, the particle number and the electron spin are conserved.  It means that some of the correlation functions are zero, and the maximum cannot be an arbitrary linear combination of all possible Hermitian operators from the subsystem, but has to be a member of one of the following five disjunct classes: spin-density correlations, charge-charge correlations, transfer-correlations, pairing, and spin-flip correlations. These are listed together with example correlators in Table \ref{tab:corr-class}. A detailed reasoning and the operators spreading the classes are given in Appendix \ref{sec:max_two_site_corr_func}.
\begin{table}[h]
    \centering
    \begin{tabular}{|c|c|c|c|}
        \hline
        \textbf{name}  & \textbf{$\Delta$S} & \textbf{$\Delta$N} & \textbf{example} \\ \hline
        Spin density  & $0$ & $0$ & $\langle s_i s_j \rangle$\\ \hline
        Charge  & $0$ & $0$ & $\langle (1-n_i)(1-n_j) ,\rangle$\\ 
                             &     &      &     $(\langle 2s^2_i-1)(2s^2_j-1)\rangle$              \\ \hline
         Transfer &  $\pm1$ &  $\pm1$ & $\langle c^{\dagger}_{i,\uparrow} c_{j,\uparrow} \rangle$\\ \hline
         Pairing & $0$&   {$\pm2$}  & $\langle c^{\dagger}_{i, \uparrow} c^{\dagger}_{i \downarrow} c_{j \uparrow} c_{j \downarrow} \rangle$\\ \hline
        Spin flip &   {$\pm2$} & $0$ & $\langle c^{\dagger}_{i \uparrow} c_{i\downarrow} c^{\dagger}_{j\uparrow} c_{j\downarrow} \rangle $\\ \hline
    \end{tabular}
    \caption{Classification of the maximal correlation function. The first column is the name of the class.  The column $\Delta S$ shows how the basis operators change the spin. The column $\Delta N$ shows how the basis operators change the particle number. The "example" column shows typical correlation functions from the class. In the example correlation functions $s_i=n_{i,\uparrow} - n_{i,\downarrow}$ and $n_i=n_{i\uparrow}+n_{i\downarrow}$.   }
    \label{tab:corr-class}
\end{table}

The nature of the maximal correlation functions is shown in Fig. \ref{Fig:max_corr_dir}. For small distances, the largest correlations are of transfer type, and with distance, there is a crossover to charge-charge type correlations. 
This corroborates our earlier discussion of entanglement and the finding that the spatial-structure is significantly shaped by the nearest-neighbor character of our model.}

\section{Summary}\label{Summary}
In this paper, we provide proof of principle that the two-site reduced density matrix can in practice be computed from Green's function based methods. 
The comparison with QMC results for the Rényi mutual information validates our approach.
This enables us to simulate systems beyond the limitations of other methods and to unveil the spatial structure of entanglement in the 2D Hubbard model. 

As regards physics, we observe a strong entanglement between NN sites in the Hubbard model and further a checkerboard-like pattern of much smaller entanglement
beyond NN sites. Entanglement is largest in the vicinity of the diagonal and suppressed along the horizontal/vertical directions. 
This diagonal pattern of entanglement can be explained by the larger number of possible hopping paths (to the same order in $t$) along the diagonal. And the checkerboard pattern is caused by the antiferromagnetic correlations
which appear already at $U=0$ for half-filling due to the perfect nesting of the Fermi surface. Upon increasing $U$, decreasing $T$ and towards half-filling,
the checkerboard entanglement pattern becomes more pronounced as do antiferromagnetic correlations.
With the observed strong entanglement between antiferromagnetic NN pairs and the checkerboard (albeit weaker) entanglement pattern 
between more distant lattice sites, our work shows that the antiferromagnetic tendency of the Hubbard model is indeed a quantum (entanglement) effect

 {Our interpretation of these results is mostly based on the characteristics given by the nearest neighbor hopping nature of our systems. An observation that is in good agreement with the dominant correlations in the system, as observed from the correlation diagnostic.}

Our observations of entanglement in the Hubbard model demonstrate that our method can provide valuable insights into the quantum nature and entanglement of strongly correlated electronic systems.
At present  $p$D\textGamma A  is restricted to smaller values of $U$ and lattices of the order of 16$\times$16. However, there are promising developments to overcome these limitations~\cite{Rohshap2025,lihm2025fdpDGA}. The presented methodology can also be applied to obtain 2s-RDM from other approximate methods that provide one- and two-particle quantities, in particular ladder extensions of DMFT~\cite{rohringer_diagrammatic_2018}. 

\begin{acknowledgements}
 {We thank Gabriele Bellomia for very fruitful discussions.} FB, FA, and KH have been supported by the SFB Q-M\&S (FWF project DOI 10.55776/F86); AK by project V 1018 of the Austrian Science Fund (FWF Grant DOI  10.55776/V1018); and GR by the QuantERA II project HQCC-101017733., National Research, Development and Innovation Office of Hungary NKFIH under Grant No.
K128989, No. K146736, and by the Hungarian Lor\'ant E\"otv\"os mobility scholarship. Calculations have been done in part on the Austrian Scientific Cluster (ASC).
\end{acknowledgements}

\section*{Data Availability}
The data that support the findings of this article are openly available \cite{bippus_2025_data}.

\appendix

\section{Entanglement in the Heisenberg dimer} \label{I_problem}
While the mutual information is a faithful entanglement measure for pure states, it can not differentiate between classical correlations and quantum entanglement for mixed states \cite{shapourian_entanglement_2019}. To illustrate this, we consider the strong coupling limit of the Hubbard model, the isotropic antiferromagnetic Heisenberg model \cite{girvin_modern_2019}
\begin{equation}
    H = \sum_{\langle i,j \rangle} J\mathbf{S}_i\mathbf{S}_j,
\end{equation} and investigate the dimer as a particularly simple example.
The density matrix in the basis $\{|\!\uparrow,\! \uparrow\rangle,|\!\uparrow, \!\downarrow\rangle,|\!\downarrow, \!\uparrow\rangle,$\\ $|\!\downarrow, \!\downarrow\rangle\}$ is
\begin{equation}
    \rho_{AB}= \frac{1}{4}\left(\begin{array}{cccc}
1+t_3 & & & t_1-t_2 \\
& 1-t_3 & t_1+t_2 & \\
& t_1+t_2 & 1-t_3 & \\
t_1-t_2 & & & 1+t_3
\end{array}\right),
\end{equation}
with $t_1=t_2=t_3 = \frac{1-\alpha^{-4}}{3+\alpha^{-4}}$ and $\alpha = e^{-\beta\frac{J}{4}}$. As long as the condition $\sum_{i=1}^{3}|t_i|\leq 1$ is fulfilled, it can be separated into the form \cite{ben-aryeh_explicit_2015}
\begin{equation}
    \begin{split}
\rho_{AB} = & \frac{1}{4}
\sum_{i=1}^3 2\left|t_i\right|\Big[\left\{\frac{\left(\mathbb{I}-\sigma_i\right)_A}{2} \otimes \frac{\left(\mathbb{I}-\operatorname{sign}\left(t_i\right) \sigma_i\right)_B}{2}\right\}\\
& +\left\{\frac{\left(\mathbb{I}+\sigma_i\right)_A}{2} \otimes \frac{\left(\mathbb{I}+\operatorname{sign}\left(t_i\right) \sigma_i\right)_B}{2}\right\}\Big] \\
&+\left[(\mathbb{I})_A \otimes(\mathbb{I})_B\right]\left(1-\sum_{i=1}^3\left|t_i\right|\right).
\end{split},
\end{equation}
Here $\sigma_i$ are the Pauli matrices. The separability condition is fulfilled for either $T=0$ or $\beta J \leq \ln 3$. However, the mutual information
\begin{equation}
    I = 2 \ln (2)+\frac{3-3 \alpha^{-4}}{3+\alpha^{-4}} \ln (\alpha)-\ln \left(3 \alpha+\alpha^{-3}\right)
\end{equation}
at $\beta J = \ln 3$ is $I=\ln\left(\frac{2}{\sqrt{3}}\right)>0$. Hence, while we know that the system is separable at $\beta J = \ln 3$, we find a non-zero mutual information measuring classical correlations and not entanglement. 
Likewise, the negativity can be computed. As we now consider a spin system, the partial transpose based negativity is applied, i.e., Eq.~\eqref{eq_negativity}. This yields
\begin{equation}
    N = \left\{ \begin{array}{l l}
    \left| \frac{-1+3\alpha^4}{2(1+3\alpha^4)}\right| & \textrm{if }\beta J \leq \ln 3 \textrm{ or } T = 0 \\
    0 & \textrm{else}
    \end{array}
    \right.,
\end{equation}
which indeed shows the correct limit of separability as it is an exact measure for qubits \cite{horodecki_separability_1996}. However, the negativity may not detect non-separable states in larger systems. To conclude, the limitation of the negativity on mutual information require a mix of both or ideally even more entanglement measures to accurately detect entangled states. 

\begin{figure*}[tp]
\includegraphics[width=\linewidth]{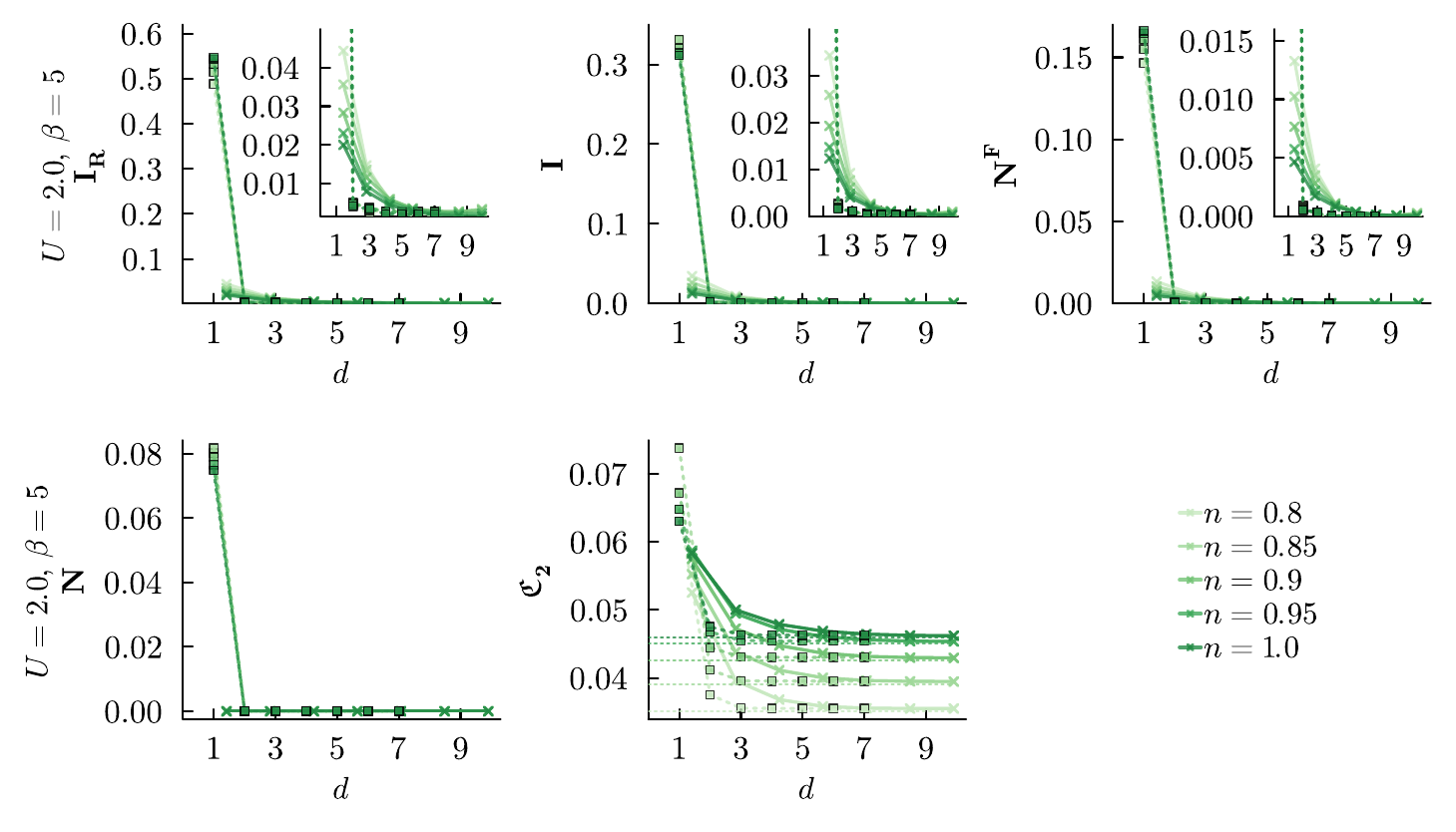}
\caption{Comparison of $I_R$, $I$, $N^F$, $N$ and $\mathfrak{C}_2$ for $U=2$ at $\beta = 5$ for a range of fillings $n$.  We show Rényi mutual information $I_{\textrm{R}}$ over distance $d$ for $U=2,\ 3,\ 4$ from left to right. Again solid lines and crosses are measurements along the diagonal of the lattice $\mathbf{\Delta}=(\Delta,\Delta)$ while boxes and dotted lines are measured along the NN path $\mathbf{\Delta}=(\Delta,0)$ and we denote the real distance $d=\sqrt{\Delta_x^2+\Delta_y^2}$ from the reference site on the $x$-axis. Insets are the same plot with an enlarged y-axis. The first row of data has already been displayed in fig.~\ref{Fig_n_Comparison}. For $\mathfrak{C}_2$ we observe that it converges towards two times the one-site limit $2\mathfrak{C}_1$ (thin constant line).
}
\label{Fig_measure_Comparison}
\end{figure*}
\begin{figure*}
\includegraphics[width=\linewidth]{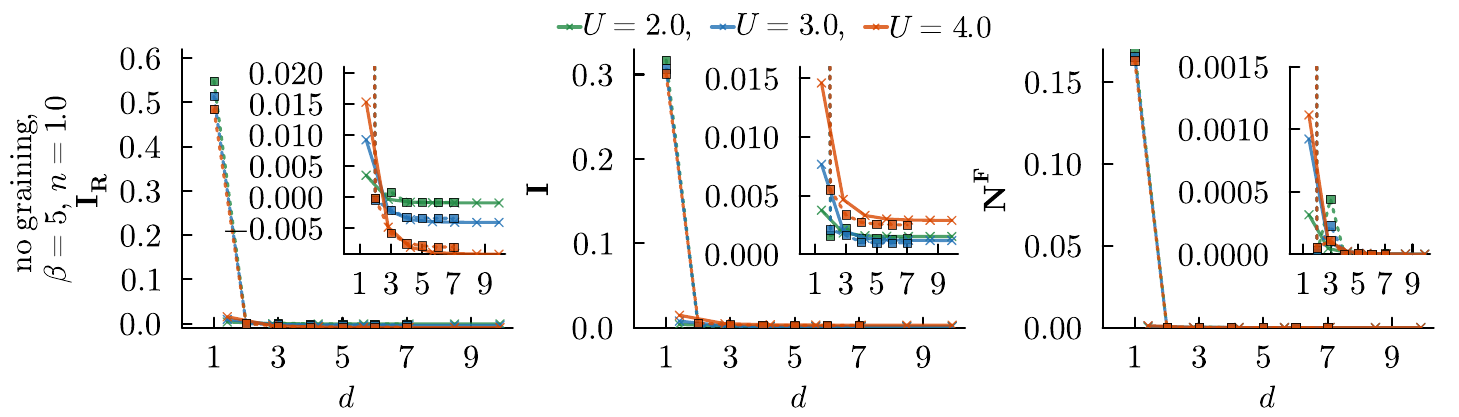}
\includegraphics[width=\linewidth]{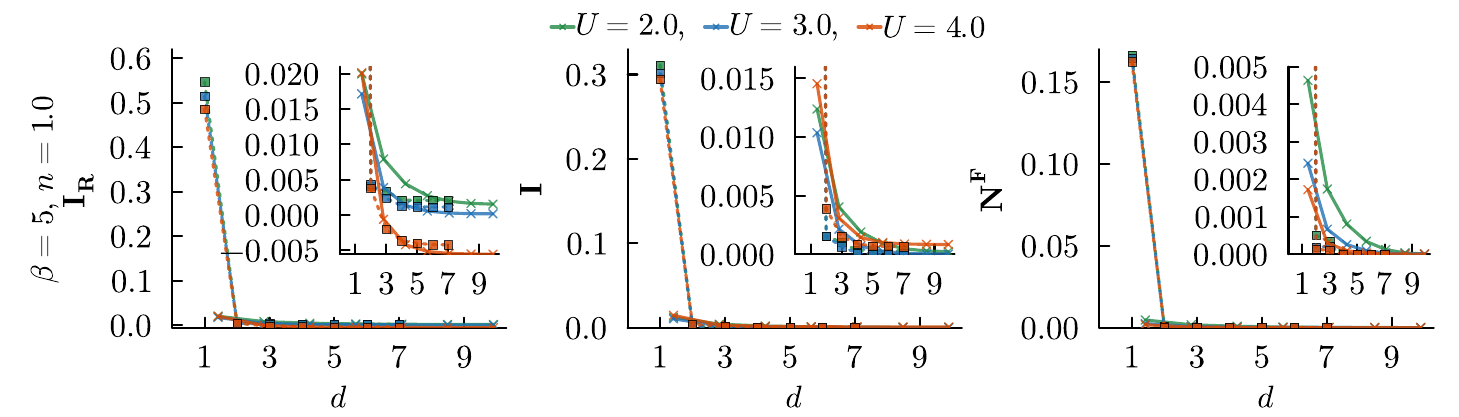}
\caption{Same as Fig.~\ref{Fig_n_Comparison} but for $U=2,\ 3,\ 4$ at $\beta = 5,\ n=1$, comparing no coarse graining (top) to coarse graining (bottom). 
}
\label{Fig_U_Comparison}
\end{figure*}

\section{Comparison of entanglement and correlation measures} \label{Measure_Comparison}
The 2s-RDM formalism allows for the computation of a wide variety of entanglement measures. In the main text, we have limited ourselves to showing results for the von Neumann and Rényi mutual information and fermionic negativity (sec.~\ref{Entanglement_measures}). Here, we want to compare these results to the plain vanilla negativity $F$ from Eq.~\eqref{eq_negativity} (without fermionic phase $\phi$) and the non-freeness \cite{held_physics_2013, gottlieb_properties_2006,bellomia2025localclassicalcorrelations}. 

First, the negativity defined through the partial transpose Eq.~\eqref{eq_negativity} does not indicate any entanglement at all beyond the NN pair, see Fig.~\ref{Fig_measure_Comparison}; it is zero. This inability to faithfully detect entanglement in fermionic systems is to be expected \cite{shapourian_partial_2017,shapourian_entanglement_2019}.
Second, the non-freeness of a density matrix $\rho$ is defined by the relative entropy from the closest free state $\Delta$
\begin{equation}
    S(\rho\|\Delta) = S(\rho) - S(\Delta),
\end{equation}
where $\mathfrak{C} = S(\rho) = - \sum_{i=1}^{\rm dim \rho} \lambda_i \ln \lambda_i$ is the von Neumann entropy of a density matrix with eigenvalues $\lambda$.
We can define two types of non-freeness. First, the non-freeness of the two site reduced density matrix $\mathfrak{C}_2$ with the closest free state given through the one that has the same one particle correlation functions as the original density matrix
\begin{equation}
    \Delta_2 =\left( \begin{array}{cccc}
        c^{\dagger}_{i\uparrow} c_{i\uparrow} & c^{\dagger}_{i\uparrow} c_{j\uparrow}   &  &\\
        c^{\dagger}_{i\uparrow} c_{j\uparrow} & c^{\dagger}_{j\uparrow} c_{j\uparrow} & & \\
     & &  c^{\dagger}_{i\downarrow} c_{i\downarrow} & c^{\dagger}_{i\downarrow} c_{j\downarrow}     \\
       & & c^{\dagger}_{i\downarrow} c_{j\downarrow} & c^{\dagger}_{j\downarrow} c_{j\downarrow} 
    \end{array} \right).
\end{equation}
Second, we can consider the distance-independent non-freeness of the one site reduced density matrix $\mathfrak{C}_1$ with free state
\begin{equation}
    \Delta_1 = \left( \begin{array}{cc}
        c^{\dagger}_{i\uparrow} c_{i\uparrow} & 0 \\
        0 & c^{\dagger}_{i\downarrow} c_{i\downarrow}
    \end{array} \right).
\end{equation}
Both matrices are directly computable from the quantities defined in Ref.~\onlinecite{roosz_two-site_2024}.
In Fig.~\ref{Fig_measure_Comparison}, we show the difference $\mathfrak{C}_2$. 
For large $d$ the result converges towards $2\mathfrak{C}_1$ (horizontal lines), showing that no correlations remain between distant sites.

To complete the comparison, the fermionic negativity and von Neumann mutual information both provide the same qualitative result and provide a clear picture of the genuine entanglement in the system. Meanwhile, the Rènyi mutual information is not well suited as it becomes negative at large distances. 


\section{The impact of coarse graining}\label{coarse_graining}
Fig.~\ref{Fig_U_Comparison} shows a direct comparison of different interaction strengths $U$
with and without coarse graining
at $n=1$ and $\beta=5$. The coarse grained data has a smoother behavior than the data without. This is to be expected as the data better simulates an infinitely sized system where finite-size effects are not relevant.


 {
\section{Two-site reduced density matrix from Green's functions}\label{2srdm_from_G_F}
In this appendix we provide the equations used to compute the 2s-RDM from two- and four-point Green's functions - which we define in Sec.~\ref{quantities}. Following Ref.~\onlinecite{roosz_two-site_2024}, the off-diagonal elements of the 2s-RDM are given by
\begin{align}
& \rho_{11,6}={\cal C}_{10}\\
& \rho_{8,9}={\cal C}_{11} \\
 &\rho_{2,4}=\rho_{3,5} =\left(\frac{\mu}{U}-1\right){\cal C}_{8A}+\frac{t}{U} {\cal C}_{12} -\frac{1}{U} {\cal C}_4 \\
 &\rho_{8,6}=\rho_{8,11}=-\rho_{9,11}=-\rho_{9,6} \nonumber\\
 &=\left (\frac{1}{2}-\frac{\mu}{U}\right){\cal C}_{8A}-\frac{1}{2}{\cal C}_{8B}-\frac{t}{U}{\cal C}_{12}+ \frac{1}{U}{\cal C}_4 \\
&\rho_{12,14}=\rho_{13,15} ={\cal C}_9-{\cal C}_{8A}-\frac{t}{U} {\cal C}_{12} +  \frac{1}{U}{\cal C}_4 - \frac{\mu}{U}{\cal C}_{8A},
\end{align}
where the  two- and four-point correlators ${\cal C}$'s are given explicitly below.
With the $\rho_{7,7}$ matrix element
\begin{align}
    &\rho_{7,7}=\rho_{10,10}= \nonumber\\
    &\frac{1}{U^2}{\cal C}_1 -\frac{t}{U^2} {\cal C}_t
-\frac{\mu}{U^2} {\cal C}_{\mu} +\frac{t^2}{U^2} {\cal C}_{t^2}
+\frac{\mu t}{U^2} {\cal C}_{\mu t}
+\frac{\mu^2}{U^2} {\cal C}_{5}
,
\end{align}
we define the following additional expectation values
\begin{align}
    \langle n_{i\uparrow} \rangle &= 1 - {\cal C}_{13}  \\
    \langle n_{i\uparrow} n_{j \uparrow} \rangle &= -1+2 \langle n_{i\uparrow} \rangle+{\cal C}_5 \\
    \langle n_{i\uparrow} n_{j \downarrow} \rangle &= -1+2 \langle n_{i\uparrow} \rangle+{\cal C}_6 \\
    \langle n_{i\uparrow} n_{i \downarrow} \rangle &= -1+2 \langle n_{i\uparrow} \rangle+{\cal C}_7 \\
    \langle n_{i\uparrow} n_{i \downarrow} n_{j \uparrow}\rangle &= -\langle n_{i\uparrow} \rangle + \langle n_{i\uparrow} n_{j \downarrow} \rangle + \langle n_{i\uparrow} n_{i \downarrow} \rangle \nonumber\\
    &+\frac{t}{U} {\cal C}_{\mu t 1}+\frac{\mu}{U} {\cal C}_5 -\frac{1}{U} {\cal C}_{\mu 2} \\
    \langle n_{i\uparrow} n_{i \downarrow} n_{j \uparrow} n_{j \downarrow} \rangle &= 
    \rho_ {7,7}+2 \langle n_{i\uparrow} n_{i \downarrow} n_{j \uparrow}\rangle - \langle n_{i\uparrow} n_{j \uparrow} \rangle.
\end{align}
Using these, we can compute the remaining diagonal elements
\begin{align}
    &\rho_{16,16}=\langle n_{i\uparrow} n_{i \downarrow} n_{j \uparrow} n_{j \downarrow} \rangle  \\
    &\rho_{1,1}= \langle n_{i\uparrow} n_{i \downarrow} n_{j \uparrow} n_{j \downarrow} \rangle -4\langle n_{i\uparrow} n_{i \downarrow} n_{j \uparrow}\rangle+2\langle n_{i\uparrow} n_{i \downarrow} \rangle \nonumber \\
    &+ 2\langle n_{i\uparrow} n_{j \uparrow} \rangle +2\langle n_{i\uparrow} n_{j \downarrow} \rangle- 4 \langle n_{i\uparrow} \rangle +1 \\ 
    &\rho_{2,2}=\rho_{3,3}=\rho_{4,4} = \rho_{5,5} \nonumber\\
    &=-\langle n_{i\uparrow} n_{i \downarrow} n_{j \uparrow} n_{j \downarrow} \rangle + 3 \langle n_{i\uparrow} n_{i \downarrow} n_{j \uparrow}\rangle - \langle n_{i\uparrow} n_{j \uparrow} \rangle\nonumber  \\
    & - \langle n_{i\uparrow} n_{j \downarrow} \rangle -\langle n_{i\uparrow} n_{i \downarrow} \rangle + \langle n_{i\uparrow} \rangle \\
    &\rho_{6,6}=\rho_{11,11} \nonumber \\
    &=\langle n_{i\uparrow} n_{i \downarrow} n_{j \uparrow} n_{j \downarrow} \rangle-2\langle n_{i\uparrow} n_{i \downarrow} n_{j \uparrow}\rangle + \langle n_{i\uparrow} n_{i \downarrow} \rangle \\
      &\rho_{8,8}=\rho_{9,9} \nonumber \\
    &=\langle n_{i\uparrow} n_{i \downarrow} n_{j \uparrow} n_{j \downarrow} \rangle-2\langle n_{i\uparrow} n_{i \downarrow} n_{j \uparrow}\rangle + \langle n_{i\uparrow} n_{j \downarrow} \rangle \\
    &\rho_{12,12}=\rho_{13,13}=\rho_{14,14}=\rho_{15,15} \nonumber \\
    &=-\langle n_{i\uparrow} n_{i \downarrow} n_{j \uparrow} n_{j \downarrow} \rangle 
    +\langle n_{i\uparrow} n_{i \downarrow} n_{j \uparrow}\rangle.
\end{align}
We now proceed by defining the ${\cal C}$ operators
\begin{align}
     {\cal C}_{5}&= (1- \langle n_{i\uparrow} \rangle)^2 \nonumber \\ & + \frac{1}{4}\sum_q (\chi_d^q+\chi_m^q) e^{i\vec{q}{(\vec{r}_i-\vec{r}_j)}},\\
      {\cal C}_{6}&=(1- \langle n_{i\uparrow} \rangle)^2 \nonumber \\ &  + \frac{1}{4}\sum_q (\chi_d^q-\chi_m^q) e^{i\vec{q}{(\vec{r}_i-\vec{r}_j)}}
     ,\\
     {\cal C}_{7} &=(1- \langle n_{i\uparrow} \rangle)^2  + \frac{1}{4}\sum_q (\chi_d^q-\chi_m^q) 
     ,\\
     {\cal C}_{10}&= \sum_q \chi_s^q \;e^{i\vec{q}{(\vec{r}_i-\vec{r}_j)}}
     ,\\
{\cal C}_{11}&=\frac{1}{2}\sum_q \chi_m^q e^{i\vec{q}{(\vec{r}_i-\vec{r}_j)}}
         ,\\
     	{\cal C}_{9}&=-\sum_k G^k e^{i\vec{k}{(\vec{r}_i-\vec{r}_j)}}
    ,\\
     	{\cal C}_{13}&= 1 - \langle n_{i\uparrow}\rangle
      ,\\
{\cal C}_{8A} &= -(1-\langle n_{i\uparrow}\rangle)\sum_{k}G^ke^{i\vec{k}{(\vec{r}_i-\vec{r}_j)}}
\nonumber \\
& - \frac{1}{2U}\sum_{k,q} \left( \gamma^{kq}_d W^q_d + \gamma^{kq}_m W^q_m\right) \nonumber \\ & \times G^kG^{k+q}e^{i\vec{k}{(\vec{r}_i-\vec{r}_j)}}
,\\
{\cal C}_{8B} &= -(1-\langle n_{i\uparrow}\rangle)\sum_{k}G^ke^{i\vec{k}{(\vec{r}_i-\vec{r}_j)}}
\nonumber \\
& - \frac{1}{2U}\sum_{k,q} \left( \gamma^{kq}_d W^q_d + \gamma^{kq}_m W^q_m\right) \nonumber \\ & \times  G^kG^{k+q}e^{i(\vec{k}+\vec{q}){(\vec{r}_i-\vec{r}_j)}}
,\\
{\cal C}_{12} &= (1-\langle n_{i\uparrow}\rangle)\frac{1}{t}\sum_{k}G^ke^{i\vec{k}{(\vec{r}_i-\vec{r}_j)}}
\varepsilon_{\vec{k}} \nonumber \\
& + \frac{1}{2tU}\sum_{k,q} \left( \gamma^{kq}_d W^q_d + \gamma^{kq}_m W^q_m\right) \nonumber \\ & \times G^kG^{k+q}e^{i\vec{k}{(\vec{r}_i-\vec{r}_j)}}
\varepsilon_{\vec{k}},\\
{\cal C}_{\mu t1} &= (1-\langle n_{i\uparrow}\rangle)\frac{1}{t}\sum_{k}G^k\varepsilon_{\vec{k}} \nonumber \\
&+\frac{1}{t}\sum_{k,q}G^kG^{k+q}e^{i\vec{q}{(\vec{r}_i-\vec{r}_j)}}
\varepsilon_{\vec{k}} \nonumber \\
& + \frac{1}{2tU}\sum_{k,q} \left( \gamma^{kq}_d W^q_d - \gamma^{kq}_m W^q_m-2U\right)\nonumber \\
& \times G^kG^{k+q}e^{i\vec{q}{(\vec{r}_i-\vec{r}_j)}}
\varepsilon_{\vec{k}},\\
{\cal C}_{\mu t2} &= (1-\langle n_{i\uparrow}\rangle)\frac{1}{t}\sum_{k}G^k\varepsilon_{\vec{k}} \nonumber \\ 
&+\frac{1}{t}\sum_{k,q}G^kG^{k+q}e^{i\vec{q}{(\vec{r}_i-\vec{r}_j)}}
\varepsilon_{\vec{k}+\vec{q}} \nonumber \\
& + \frac{1}{2tU}\sum_{k,q} \left( \gamma^{kq}_d W^q_d - \gamma^{kq}_m W^q_m - 2U\right)\nonumber \\
&\times G^kG^{k+q}e^{i\vec{q}{(\vec{r}_i-\vec{r}_j)}}
\varepsilon_{\vec{k}+\vec{q}},
,\\
{\cal C}_{4} &= (1-\langle n_{i\uparrow}\rangle)\sum_{k}i\nu(G^k - \frac{1}{i\nu})e^{i\vec{k}{(\vec{r}_i-\vec{r}_j)}}
 \nonumber \\
& + \frac{1}{2U}\sum_{k,q}i\nu \left( \gamma^{kq}_d W^q_d + \gamma^{kq}_m W^q_m\right) \nonumber \\ & \times G^kG^{k+q}e^{i\vec{k}{(\vec{r}_i-\vec{r}_j)}},\\
{\cal C}_{\mu 1} &= (1-\langle n_{i\uparrow}\rangle)\sum_{k}i\nu(G^k - \frac{1}{i\nu})  \nonumber \\
& +\sum_{k,q}i(\nu+\omega)\left(G^{k+q} - \frac{1}{i(\nu+\omega)}\right)G^{k}e^{i\vec{q}{(\vec{r}_i-\vec{r}_j)}} \nonumber \\
& +  \frac{1}{2U}\sum_{k,q}i(\nu+\omega) \left(\gamma^{kq}_d W^q_d - \gamma^{kq}_m W^q_m -2U\right) \nonumber \\
&\times G^kG^{k+q}e^{i\vec{q}{(\vec{r}_i-\vec{r}_j)}},\\
{\cal C}_{\mu 2} &= (1-\langle n_{i\uparrow}\rangle)\sum_{k}i\nu(G^k - \frac{1}{i\nu}) \nonumber \\ 
& +\sum_{k,q}i\nu(G^k - \frac{1}{i\nu})G^{k+q}e^{i\vec{q}{(\vec{r}_i-\vec{r}_j)}}
 \nonumber \\
 & + \frac{1}{2U}\sum_{k,q}i\nu \left( \gamma^{kq}_d W^q_d - \gamma^{kq}_m W^q_m -2U\right) \nonumber\\ & \times G^kG^{k+q}e^{i\vec{q}{(\vec{r}_i-\vec{r}_j)}},\\
	 &{\cal C}_1 = -\sum_{k}\nu\left( G^k - \frac{1}{i\nu} \right)\sum_{k'}\nu'\left(G^{k'} - \frac{1}{i\nu'} \right) \nonumber \\
  &+\sum_{k,q}\nu(\nu+\omega)\left(G^k-  \frac{1}{i\nu}\right) \nonumber \\ & \times \left(G^{k+q} - \frac{1}{i(\nu+\omega)} \right)e^{i\vec{q}{(\vec{r}_i-\vec{r}_j)}} \nonumber \\
  &+ \frac{1}{2}\sum_{k,k',q}\nu(\nu'+\omega)G^k G^{k+q}G^{k'}G^{k'+q} \nonumber \\
  &\times (F_d^{kk'q}+ F_m^{kk'q})e^{i\vec{q}{(\vec{r}_i-\vec{r}_j)}},\\
	 &{\cal C}_{t1} = \frac{1}{t}\sum_{k} \varepsilon_{\vec{k}}  G^k \sum_{k'}i\nu'\left(G^{k'} - \frac{1}{i\nu'} \right)\nonumber \\
  &- \frac{1}{t}\sum_{k,q}i(\nu+\omega)G^k \left(G^{k+q} - \frac{1}{i(\nu+\omega)} \right) \nonumber \\ & \times e^{i\vec{q}{(\vec{r}_i-\vec{r}_j)}}\varepsilon_{\vec{k}}   \nonumber \\
  &- \frac{1}{2t}\sum_{k,k',q}i(\nu'+\omega)G^k G^{k+q}G^{k'}G^{k'+q} \nonumber \\
  &\times (F_d^{kk'q}+ F_m^{kk'q})e^{i\vec{q}{(\vec{r}_i-\vec{r}_j)}}\varepsilon_{\vec{k}},\\
	 &{\cal C}_{t2} = \frac{1}{t}\sum_{k}i\nu\left(G^{k} - \frac{1}{i\nu} \right)  \sum_{k'}\varepsilon_{\vec{k'}}G^{k'} \nonumber \\
  &- \frac{1}{t}\sum_{k,q}i\nu\left(G^{k} - \frac{1}{i\nu} \right)G^{k+q} e^{i\vec{q}{(\vec{r}_i-\vec{r}_j)}}\varepsilon_{\vec{k}+\vec{q}}   \nonumber \\
  &- \frac{1}{2t}\sum_{k,k',q}i\nu G^k G^{k+q}G^{k'}G^{k'+q} \nonumber \\
  &\times (F_d^{kk'q}+ F_m^{kk'q})e^{i\vec{q}{(\vec{r}_i-\vec{r}_j)}}\varepsilon_{\vec{k}'+\vec{q}},\\
	 &{\cal C}_{t^2} = \frac{1}{t^2}\sum_{k}\varepsilon_{\vec{k}}G^{k} \sum_{k'}\varepsilon_{\vec{k'}}G^{k'} \nonumber \\
  &- \frac{1}{t^2}\sum_{k,q}G^{k}G^{k+q} e^{i\vec{q}{(\vec{r}_i-\vec{r}_j)}}\varepsilon_{\vec{k}}\varepsilon_{\vec{k}+\vec{q}}   \nonumber \\
  &- \frac{1}{2t^2}\sum_{k,k',q}G^k G^{k+q}G^{k'}G^{k'+q} \nonumber \\
  &\times (F_d^{kk'q}+ F_m^{kk'q})e^{i\vec{q}{(\vec{r}_i-\vec{r}_j)}}\varepsilon_{\vec{k}}\varepsilon_{\vec{k}'+\vec{q}}.
\end{align}
}

\section{Comparison of susceptibilities}\label{comp_chi}
\begin{figure*}[tp]
\includegraphics[width=\linewidth]{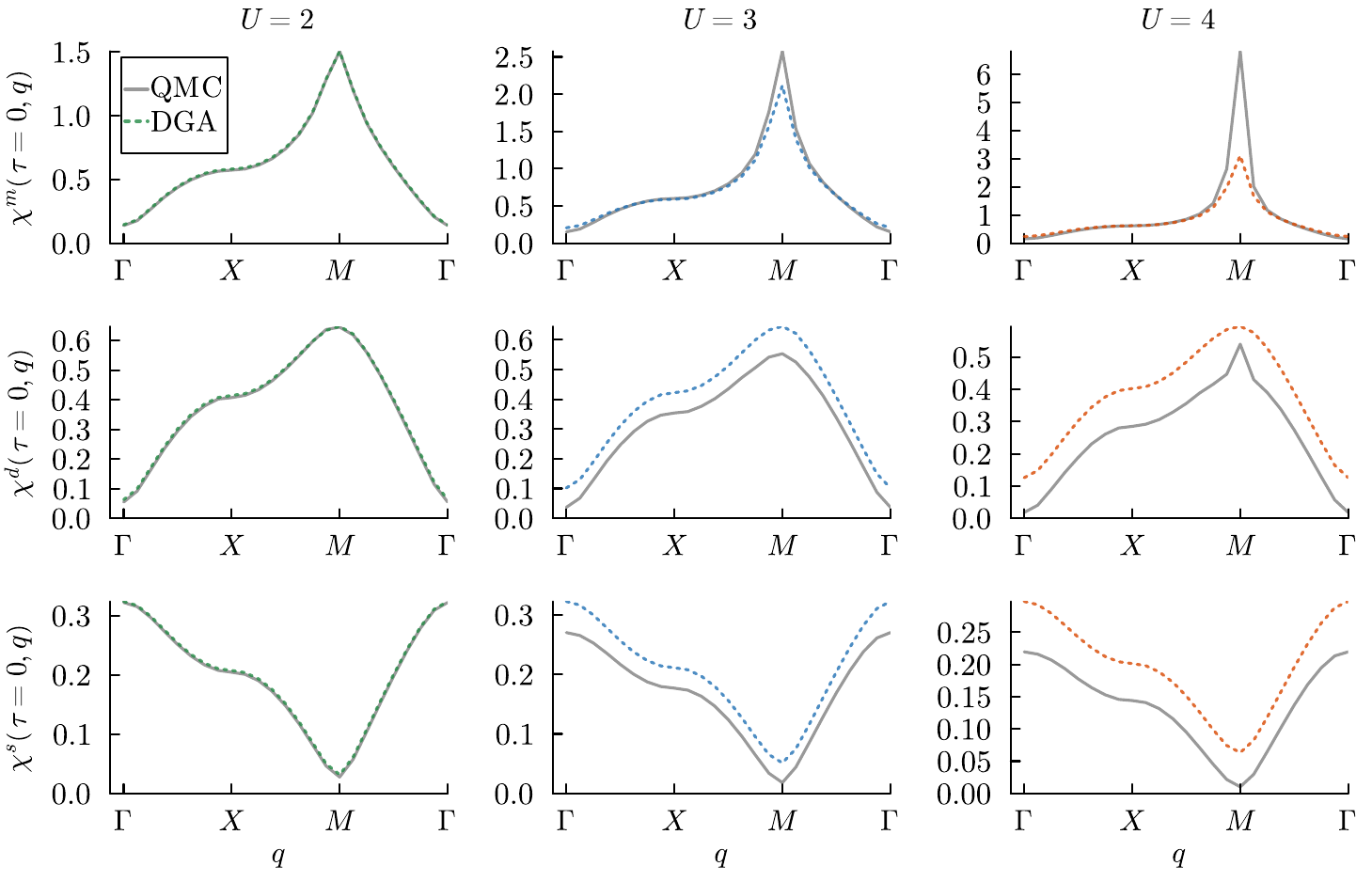}
\caption{Comparison of susceptibilities $\chi(\tau = 0, q)$ along high symmetry path $\Gamma-X-M-\Gamma$ obtained with $p$D\textGamma A without coarse-graining (dotted line in color) to QMC (solid grey line). For a) $U=2$, b) $U=3,$, c) $U=4$ at $\beta = 5,\ n=1$.
}
\label{Fig_QMC_Comparison_chi_q}
\end{figure*}

In the calculation of the 2s-RDM the spin, density, and singlet susceptibilities of Eq.~\eqref{eq_chi} contribute~\cite{roosz_two-site_2024}. Since these can be directly compared between $p$D\textGamma A and QMC they provide an excellent benchmark. We note that the definition of Eq.~\eqref{eq_chi} corresponds to the following imaginary time correlators 
\begin{align}
&\chi^q_m = \sum_{j} e^{-i\mathbf{q}{(\mathbf{r}_i-\mathbf{r}_j)}}\langle S^z_i(\tau)S^z_j\rangle , \\ 
&\chi^q_d = \sum_{j} e^{-i\mathbf{q}{(\mathbf{r}_i-\mathbf{r}_j)}}\langle n_i(\tau)n_j\rangle -\beta \delta_q \langle n_i\rangle^2 ,
\\ 
&\chi^q_s = \sum_{j}e^{-i\mathbf{q}{(\mathbf{r}_i-\mathbf{r}_j)}}\langle c_{i\downarrow}(\tau)c_{i\uparrow}(\tau)c^\dagger_{j\uparrow}c^\dagger_{j\downarrow}\rangle\kh{.} 
\end{align}
We benchmark them as a function of $\mathbf{k}$ along the high symmetry path $\Gamma-X-M-\Gamma$ through the Brillouin zone at imaginary time $\tau = 0$ in  Fig.~\ref{Fig_QMC_Comparison_chi_q}. Excellent agreement is observed at $U=2$, which for the same parameters was also shown in Refs.~\onlinecite{Hille2020, SchaeferPRX} (where also parquet approximation, valid for $U=2$ was applied). At  $U=4$ the agreement is only qualitative. The susceptibility at the $M$ point is largely enhanced in QMC, a trend not yet observed in $p$D\textGamma A at this $T$ and $U$ as explained in the main text. 

\begin{figure}[tp]
\includegraphics[width=\linewidth]{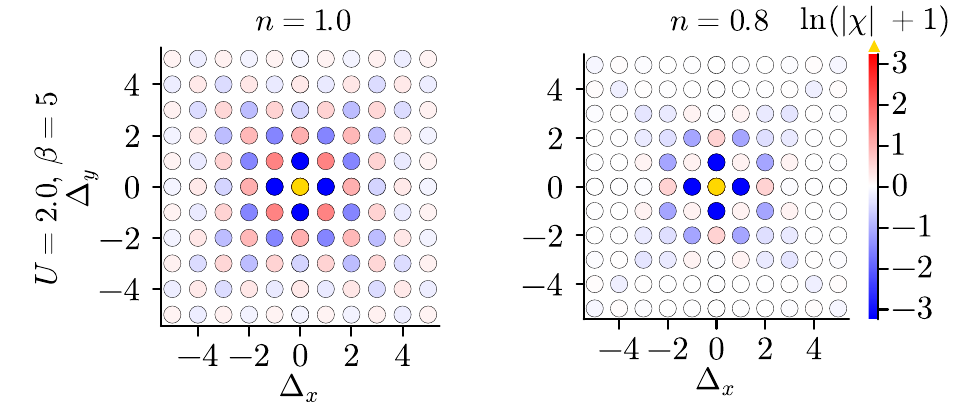}
\caption{Susceptibilities at $U=2$, $\beta = 5$ for $n=1.0$ and $n=0.8$. To best visualize the entire data range, we plot $\textrm{sgn}[\chi^m(\tau=0,\mathbf{r})]\ln[|\chi^m(\tau=0,\mathbf{r})|+1]$.
}
\label{fig_susc}
\end{figure}

In the context of susceptibilities, we also show the $\chi^{m}(\tau=0,\mathbf{r})$ susceptibility for $n=1.0$ and $n=0.8$ at $U=2$, $\beta=5$ in Fig.~\ref{fig_susc}. This shows the antiferromagnetic Neel-like structure at half-filling which is 
suppressed off half-filling.

\section{Mutual information in the non-interacting case}\label{Mutualinformation_U0}
In this section, we discuss the two-point mutual information in the two-dimensional non-interacting lattice, at zero and finite temperatures.  In this case, both the density matrix of the whole system and the reduced density matrices are fermionic Gaussian matrices and standard techniques can be used \cite{Peschel_2009}.
One only needs the occupation $\langle c^\dagger_{i,\sigma}  c_{i,\sigma}\rangle$ and the hopping expectation value $\langle c^\dagger_{i,\sigma}  c_{j,\sigma}\rangle$ to reconstruct the two site density matrix, and these values can be obtained as 
\begin{equation}
    \langle c^\dagger_{i,\sigma}  c_{j,\sigma}\rangle = \frac{1}{L^2}\sum_k n_k e^{i k_x d_x +  i k_y d_y}  
    \label{eq:hopping}
\end{equation}
where $n_k=1/(\exp(\beta(\epsilon_k-\mu)) +1)$ is the fermion occupation number, and $(d_x,d_y)$ is the vector connecting sites $i$ and $j$.
In the special case of zero temperature, half\kh{-}filling, and in the infinite size limit, one can perform the sum in E.q. (\ref{eq:hopping}) 
\begin{equation}
 \langle c^{\dagger}_i c_j \rangle  = \begin{cases}
           0  \textnormal{ \,, \, if \, }   (i_x-j_x)^2-(i_y-j_y)^2=0   \\
           \\
           \displaystyle{-\frac{(-1)^{i_x-j_x}-(-1)^{i_y-j_y}}{\pi^2 \left[ (i_x-j_x)^2 - (i_y-j_y)^2 \right]}}  \textnormal{, otherwise}   \\
          \end{cases}\;.
\label{eq:U0T0mu0_hopping}
\end{equation}
In the general case, one computes Eq.~(\ref{eq:hopping}) numerically and gets the mutual information as
\begin{align}
I &= S_i + S_j - S_{i\cup j}\\
&= 4 \ln \langle c^\dagger_{i,\uparrow}  c_{i,\uparrow}\rangle \\
&-  4 (\langle c^\dagger_{i,\uparrow}  c_{i,\uparrow}\rangle   \!- \!\langle c^\dagger_{i,\uparrow}  c_{j,\uparrow}\rangle) \ln (\langle c^\dagger_{i,\uparrow}  c_{i,\uparrow} \rangle \!-\! \langle c^\dagger_{i,\uparrow}  c_{i,\uparrow}\rangle)   \\
& -  4 (\langle c^\dagger_{i,\uparrow}  c_{i,\uparrow}\rangle  \!+\! \langle c^\dagger_{i,\uparrow}  c_{j,\uparrow}\rangle) \ln (\langle c^\dagger_{i,\uparrow}  c_{i,\uparrow} \rangle \!+\! \langle c^\dagger_{i,\uparrow}  c_{i,\uparrow}\rangle) 
\label{eq:non-int-muti}
\end{align}

\begin{figure}[tb]
\includegraphics[width=\linewidth]{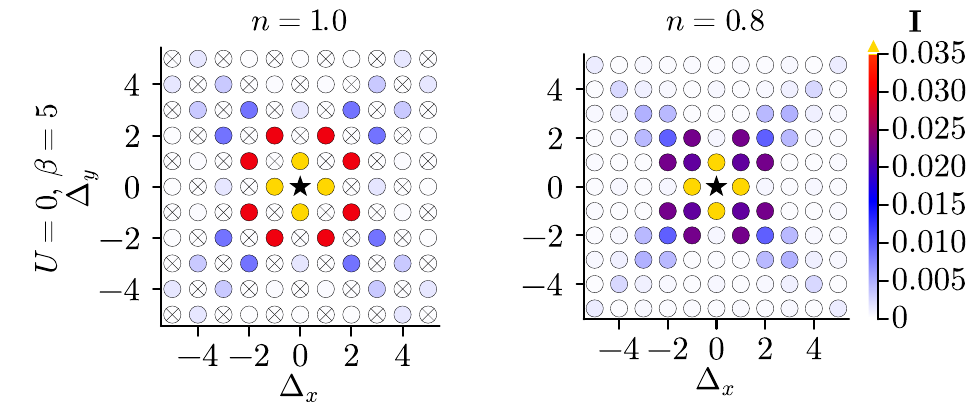}
\caption{Left panel: Mutual information in a $1024 \times 1024$ non-interacting lattice ($U=0$), at half-filling and $\beta=5$. The crosses denote sites where the mutual information is exactly zero, the colorful circles denote the magnitude of the mutual information at the sites where it is non-zero, according to the scale to the right of the figure. Right panel: same as left but at $n=0.8$.}
\label{Fig_non_int_half_fil_muti}
\end{figure}

In the ground state of the half-filled infinite non-interacting system for large distances, this yields
\begin{equation}
I \approx  \begin{cases}
           0  \textnormal{, if  }  \quad (i_x-j_x)^2-(i_y-j_y)^2=0   \\
           \\
           \displaystyle{ 8 \left\{  \frac{(-1)^{i_x-j_x}-(-1)^{i_y-j_y}}{\pi^2 \left[ (i_x-j_x)^2 - (i_y-j_y)^2 \right]} \right\}^2 }  \textnormal{, otherwise}   \\
          \end{cases}\;.
\label{eq:I_ground_state}
\end{equation}
Here we expanded the $x \ln x+(1-x) \ln (1-x)$ function appearing in Eq.~(\ref{eq:non-int-muti}) in a Taylor series up to second order around its maximum at  $x=1/2$, and in this way obtain a formula which is asymptotically exact for large distances. Numerical comparison shows that the formula has less than 5\% error even when used for the nearest neighbors.
Eq. (\ref{eq:I_ground_state}) means, that the two point mutual information decreases with the fourth power of the distance ($I \sim 1/d^4$) at zero temperature, and there is a checkerboard structure in $I$. This structure survives at finite temperatures, as it is shown in Fig. \ref{Fig_non_int_half_fil_muti}.
If the system is not half-filled, there are no exactly vanishing correlations, and no checkerboard structure, as it is shown in the right panel of Fig. \ref{Fig_non_int_half_fil_muti}.

\section{Correlation diagnostics in a general system}\label{sec:max_corr_function}
 {
We consider two distinct subsystems of a physical system $A$ and $B$.
The dimensions of the corresponding Hilbert spaces are $d_A$ and $d_B$. The system $A \cup B$ is described by the density matrix $\rho$. $O_A$ is an   operator located in $A$ and $O_B$ is an operator located in $B$. Let's assume that the operators are normalized 
\begin{equation}
	Tr O^2_A = Tr O^2_B =1.
\end{equation}
We want to find the operators maximizing the correlation function: 
\begin{equation}
\cal K(O_A,O_B)=\langle O_A \rangle  \langle O_B \rangle - \langle O_A O_B \rangle.  	
\end{equation}

The key observation to find the maximum is that $\cal{K}$ is linear in $O_A$ and $O_B$. Therefore, if we define two bases on the operators of $H_A$ and $H_B$, we can construct a matrix  $\cal{K}$, which represents the correlation functions. (This matrix is the function of the physical state $\rho$).

We set a basis on the space of operators for subsystem $A$ and subsystem $B$
\begin{align}
	&b^{(A)}_n  \quad n=1 \dots d^2_A\\
	&Tr({b^A_n}^\dagger b^A_m ) =\delta_{n,m}
\end{align}
	
\begin{align}
	&b^{(B)}_n  \quad n=1 \dots d^2_B\\
	&Tr({b^B_n}^\dagger b^B_m ) =\delta_{n,m}
\end{align}
Expansion of a given operator in the basis is then
\begin{align}
	O_A = \sum_{n=1}^{d^2_A} Tr(b^{(A) \dagger}_n O_A ) b^{(A)}_n = \sum_{n=1}^{d^2_A} \alpha^{(A)}_n  b^{(A)}_n \\
		O_B = \sum_{n=1}^{d^2_B} Tr( b^{(B) \dagger}_n O_B ) b^{(B)}_n = \sum_{n=1}^{d^2_B} \alpha^{(B)}_n  b^{(B)}_n.
\end{align}
Here $\alpha^{(B)}_n$ and $\alpha^{(A)}_n$ are complex numbers, the expansion coefficients.
The matrix defining the correlations is a $d^2_A \times d^2_B$ matrix

\begin{align}
	M_{n,m} =\langle b^{(A)}_n b^{(B)}_m \rangle - \langle b^{(A)}_n \rangle  \langle b^{(B)}_m \rangle . \label{eq:M_with_corr}
\end{align}
E.q. (\ref{eq:M_with_corr}) can be evaluated using the reduced density matrix of $A\cup B$ system
\begin{equation}
M_{n,m} =Tr[ \rho_{A \cup B} b^{(A)}_n b^{(B)}_m ]- Tr [ \rho_{A \cup B} b^{(A)}_n ]   Tr [ \rho_{A \cup B} b^{(B)}_m ] . \label{eq:M_with_corr_Tr}    
\end{equation}

The value of a correlation function can be written as
\begin{equation}
{ \cal K} (O_A,O_B) =\sum_{n=1}^{d^2_A} \sum_{m=1}^{d^2_B}  \alpha^A_n M_{n,m} \alpha^B_m.	
\end{equation}
To find the pair $(O_A, O_B)$ which maximizes the correlation function one can consider the singular value decomposition of $M$
\begin{equation}
	M=U \Sigma V^*
\end{equation} 
and the left/right vectors corresponding to the maximal singular value give the operators maximizing the correlation function.}

\section{ Maximal two-site correlation functions in the Hubbard model}\label{sec:max_two_site_corr_func}
 {
Now we consider the special case where one subsystem is a given site in the Hubbard model "i", and the other subsystem is another site denoted by "j".
 The local Hilbert space of a given site is 4 dimensional, therefore the linear space of operators acting on a given site is $4^2=16$ dimensional. 

 Due to the particle number and spin symmetries, and the translational invariance of the model, the $M$ matrix Eq.~(\ref{eq:M-matrix}) of the correlation functions splits into subspaces.

 There is no correlation between operators from different subspaces. The operator corresponding to the maximal correlation function can be only a combination of operators from one subspace.
 One can characterize the nature of the maximum correlation function, by giving its subspace. 
Based on the subspace structure we use five classes for the maximum correlations:  spin-spin, density-density, charge transfer, pairing, and spin-flip. Some of these categories correspond to a given subspace, and some of them include two subspaces with similar properties. 

Here, we list the operators used as a basis. There is a basis for the operators of site "$i$", $b^{(i)}_n$ ($n=1 \dots 16$) and a similar basis for site "$j$", $b^{(j)}_n$ ($n=1 \dots 16$).  These are selected to be orthogonal under the trace operation, and normalized. While we used a Hermitian basis, it is not required. 
We list here the basis for the $i$ site; the basis of the "$j$" site is identical, 
just defined on the $j$ site.
We list the basis operators, and define the subspaces in the same time:
\begin{equation}
b^{(i)}_1=1/2 
\end{equation}
Unit operator, not correlated with other operators

The {\it spin-spin correlation class} is defined  by $b^{(i)}_2$
\begin{equation}
	b^{(i)}_2=\frac{1}{\sqrt{2}}(n_{i,\uparrow} - n_{i,\downarrow})
\end{equation}

 The {\it  charge density-density correlation class}  is spanned by the operators  $b_3$ and $b_4$
\begin{align}
b^{(i)}_3&=~\frac{1}{\sqrt{2}}(1-n_{i,\uparrow} -n_{i,\downarrow}) \\
b^{(i)}_4&=\frac{1}{2} (2(n_{i\uparrow}-n_{i\downarrow})^2-1)
\end{align}

The {\it electron transfer correlation class} with spin up contains correlations between $b^{(i)}_5 \dots b^{(i)}_8$ and
$b^{(j)}_5 \dots b^{(j)}_8$:
\begin{align}
	 b^{(i)}_5 &= \frac{1}{\sqrt{2}}[c_{i,\uparrow}^\dagger  (1-n_{i,\downarrow}) +  (1-n_{i,\downarrow})  c_{i,\uparrow}] \\ 
	 b^{(i)}_{6} &=\frac{1}{\sqrt{2}}[ c_{i,\uparrow}^\dagger  n_{i,\downarrow} +   n_{i,\downarrow} c_{i,\uparrow} ] \\
	  b^{(i)}_{7} &=\frac{i}{\sqrt{2}}[c_{i,\uparrow}^\dagger  (1-n_{i,\downarrow}) -  (1-n_{i,\downarrow})  c_{i,\uparrow}] \\
	  b^{(i)}_{8} &= \frac{i}{\sqrt{2}}[ c_{i,\uparrow}^\dagger  n_{i,\downarrow} -   n_{i,\downarrow} c_{i,\uparrow} ]
\end{align}

The {\it transfer class with spin down}  is
\begin{align}
	b^{(i)}_9 & = \frac{1}{\sqrt{2}}[c_{i,\downarrow}^\dagger (1-n_{i,\uparrow})  + (1-n_{i,\uparrow})  c_{i,\downarrow} ] \\
    b^{(i)}_{10} & =\frac{1}{\sqrt{2}}[c_{i,\downarrow}^\dagger n_{i,\uparrow}  + n_{i,\uparrow}  c_{i,\downarrow} ] \\
   	 b^{(i)}_{11} & = \frac{i}{\sqrt{2}}[c_{i,\downarrow}^\dagger (1-n_{i,\uparrow})  - (1-n_{i,\uparrow})  c_{i,\downarrow} ] \\
   	 b^{(i)}_{12} &= \frac{i}{\sqrt{2}}[c_{i,\downarrow}^\dagger n_{i,\uparrow}  - n_{i,\uparrow}  c_{i,\downarrow} ]\; .
\end{align}

The {\it  pairing correlation, symmetric class} is

\begin{equation}
   b^{(i)}_{13} = \frac{1}{\sqrt{2}}[ c_{i,\uparrow}^\dagger c_{i,\downarrow}^\dagger  +  c_{i,\downarrow} c_{i,\uparrow} ]\; .
\end{equation}

The {\it  pairing correlation, antisymmetric class} is

\begin{equation}
	 b^{(i)}_{14} = \frac{i}{\sqrt{2}}[ c_{i,\uparrow}^\dagger c_{i,\downarrow}^\dagger  -  c_{i,\downarrow} c_{i,\uparrow} ]\; .
\end{equation}

The {\it  spin flip, symmetric class} is
\begin{equation}
	b^{(i)}_{15} = \frac{1}{\sqrt{2}}[ c_{i,\uparrow} c_{i,\downarrow}^\dagger  + c_{i,\downarrow} c_{i,\uparrow}^\dagger]\; .
\end{equation}

{\it  Spin flip, antisymmetric class} is

\begin{equation}
	 b^{(i)}_{16} = \frac{i}{\sqrt{2}}[ c_{i,\uparrow} c_{i,\downarrow}^\dagger  -   c_{i,\downarrow} c_{i,\uparrow}^\dagger]\; .
\end{equation}

\begin{equation}
{ \cal K} (O_i,O_j) =\sum_{n=1}^{16} \sum_{m=1}^{16}  \alpha^i_n M_{n,m} \alpha^j_m	
\end{equation}
The matrix elements of the $M$ matrix are quadratic polynomials of the matrix elements of the $\rho_{ij}$ density matrix given by Eq.~\ref{eq:M_with_corr_Tr}. For example 
\begin{align}
    &M_{2,2} = \frac{1}{2}(\rho_{10,10}-\rho_{9,9}-\rho_{8,8}+\rho_{7,7})\\
    &-\frac{1}{2}\left[\rho_{8,8}-\rho_{4,4}-\rho_{5,5}-\rho_{7,7}\nonumber \right. \\
    &\left. -\rho_{9,9}-\rho_{10,10}-\rho_{12,12} +\rho_{13,13} \right] \left[ \rho_{3,3}-\rho_{2,2}\right.\nonumber \\
    &  \left. -\rho_{7,7}-\rho_{8,8}+\rho_{9,9}+\rho_{10,10}-\rho_{14,14}+\rho_{15,15} \right] \nonumber 
\end{align}
}
\begin{widetext}
 \begin{equation}
 \label{eq:M-matrix}
M =	\left[\begin{array}{cccccccccccccccc}
		0 &  \rtemp   & \multicolumn{13}{l}{ \leftarrow  \textrm{This 0 corresponds to the unit operator} }  \\ \cline{1-2}
		  \ltemp   & M_{2,2} &  \rtemp  &   \multicolumn{12}{l}{ \leftarrow  \textrm{Spin - Spin correlation class }} \\ \cline{2-4}
		    &    \ltemp & M_{3,3} & M_{3,4} & \rtemp & \multicolumn{10}{l}{ \leftarrow  \textrm{Density-density correlations  }} \\
		    &    \ltemp  & M_{4,3} & M_{4,4} & \rtemp & & & & & & & & & & & \\ \cline{3-8}
                   &    &  & \ltemp & 0 & M_{5,6} & M_{5,7}  & M_{5,8} &\rtemp & \multicolumn{6}{l}{ \leftarrow  \textrm{Transfer with spin up  }} \\
                   &    &  & \ltemp & M_{6,5} & 0 & M_{6,7} & M_{7,8} & \rtemp & & & & & & & \\       
              &  &   &    \ltemp & M_{7,5}  &  M_{7,6} &  0 & 	0  & \rtemp& & & & & & & \\
              &  &   &     \ltemp & M_{8,5}  & M_{8,6} & 0 & 0  & \rtemp& & & & & & & \\ \cline{5-12}         
             \multicolumn{7}{r}{ \textrm{Transfer with spin down  } \rightarrow} & \ltemp & 0 & M_{9,10} & M_{9,11} & M_{9,12}&   \rtemp & & & \\ 
            & &  &  &   &    &  &  \ltemp & M_{10,9} & 0   &  M_{10,11} & M_{10,12}& \rtemp& & & \\           
            &  &   &  & &  &  & \ltemp & M_{11,9} & M_{11,10} & 0 & 0    & \rtemp & & & \\
            &  &   &  & &  &  & \ltemp & M_{12,9} & M_{12,10} & 0 & 0    & \rtemp & & & \\ \cline{9-13}           
            \multicolumn{11}{r}{ \textrm{Pairing, symmetric } \rightarrow} & \ltemp &M_{13,13} &  \rtemp  &  &  \\ \cline{13-14}
              \multicolumn{12}{r}{ \textrm{Pairing, antisymmetric } \rightarrow} & \ltemp& M_{14,14}    & \rtemp &  \\ \cline{14-15}         
              \multicolumn{13}{r}{ \textrm{Spin-flip, symmetric } \rightarrow} &  \ltemp   & M_{15,15} &  \rtemp\\ \cline{15-16}
               \multicolumn{14}{r}{ \textrm{Spin-flip, antisymmetric } \rightarrow }&   \ltemp   &  M_{16,16} 
	\end{array}\right] .
\end{equation}

\end{widetext}

\bibliography{references.bib}

\begin{thebibliography}{96}
\bibitem [{foo({\natexlab{b}})}]{footnoteN}%
  \BibitemOpen
  \href@noop {} {} ({\natexlab{b}}),\ \bibinfo {note} {in the literature, the negativity is also often considered as an upper bound for entanglement. In these cases, its property to bound distillable entanglement from above is referenced \cite{wang_logarithmic_2020}.}\BibitemShut {Stop}%
\bibitem [{foo({\natexlab{c}})}]{footnoteRS}%
  \BibitemOpen
  \href@noop {} {} ({\natexlab{c}}),\ \bibinfo {note} {note that only the symmetry-reduced wedge $(\Delta_x,\Delta_y),\ \Delta_x>0,\ \Delta_x\geq \Delta_y$ has been directly computed in $p$D\textGamma A and the remaining space has been filled by symmetry relations.}\BibitemShut {Stop}%
\end{thebibliography}

\end{document}